\documentclass[a4paper,11pt]{article}
\pdfoutput=1 

\usepackage{jcappub} 
\usepackage{subfig}
\usepackage[T1]{fontenc} 

\title{\boldmath Radial stability of anisotropic strange quark stars}
\author{Jos\'e D.V. Arba\~nil,$^{1,2}$}
\author{M. Malheiro$^1$}
\affiliation{$^1$ITA - Instituto Tecnol\'ogico de Aeron\'autica - Departamento de F\'isica, 12228-900, S\~ao Jos\'e dos Campos, S\~ao Paulo, Brazil}
\affiliation{$^2$UPN - Universidad Privada del Norte - Departamento de Ciencias, Av. Alfredo Mendiola 6062 Urb. Los Olivos, Lima, Lima, Per\'u}
\emailAdd{jose.arbanil@upn.pe}
\emailAdd{malheiro@ita.br}
\abstract{The influence of the anisotropy in the equilibrium and stability of strange stars is investigated through the numerical solution of the hydrostatic equilibrium equation and the radial oscillation equation, both modified from their original version to include this effect. The strange matter inside the quark stars is described by the MIT bag model equation of state. For the anisotropy two different kinds of local anisotropic $\sigma=p_t-p_r$ are considered, where $p_t$ and $p_r$ are respectively the tangential and the radial pressure: one that is null at the star's surface defined by $p_r(R)=0$, and one that is nonnull at the surface, namely, $\sigma_s=0$ and $\sigma_s\neq0$. In the case $\sigma_s=0$, the maximum mass value and the zero frequency of oscillation are found at the same central energy density, indicating that the maximum mass marks the onset of the instability. For the case $\sigma_s\neq0$, we show that the maximum mass point and the zero frequency of oscillation coincide in the same central energy density value only in a sequence of equilibrium configurations with the same value of $\sigma_s$. Thus, the stability star regions are determined always by the condition $dM/d\rho_c>0$ only when the tangential pressure is maintained fixed at the star surface's $p_t(R)$. These results are also quite important to analyze the stability of other anisotropic compact objects such as neutron stars, boson stars and gravastars.}

\begin{document}
\maketitle
\flushbottom

\section{Introduction}\label{sec-introd}

In astrophysics, the study of compact stars is vitally important because it provides an excellent laboratory for the study of highly dense matter on extreme conditions. Theoretically, it is largely considered that these objects are composed of a perfect fluid. However, strong theoretical evidences suggest that in a highly dense fluid, as the one contained in a compact star, different physical phenomena could give rise to anisotropy. For example, because of the geometry of the $\pi^{-}$ modes, anisotropic distribution of pressure could be considered in a neutron star to describe a pion condensed phase configuration \cite{sawyer}. Anisotropy could be caused within a neutron star due to the presence of a solid or a superfluid core \cite{ruderman1972,heiselberg2000}. It appears in self-gravitating objects determined by complex scalar fields, i.e., boson stars, \cite{gleiser1988,gleiser1989}. In addition, the anisotropy could be of importance in other context, for instance, to explore the hypothesis that the event GW$150914$ \cite{Gravi_waves} could be produced by the merging of rotating gravastars and not by black holes \cite{chirenti_rezzolla2016}. Independently of the nature of the anisotropy, it might produce considerable changes in the physical properties of compact stars.

In the scope of General Relativity we can find a vast number of works studying the influence of the anisotropy on static spherically symmetric objects composed of a perfect fluid (review \cite{mak_harko2003,herrera_santos1997} and references therein). In order to see the influence of the anisotropy on physical properties of spherically symmetric objects made of an isotropic fluid, the anisotropic stress tensor is added to the perfect fluid energy-momentum tensor. The components of the resulting energy-momentum tensor $T^{1}_{1}$ and $T^{2}_{2}=T^{3}_{3}$ are usually renamed as the radial and tangential pressure, $p_r$ and $p_t$ respectively. The difference between the pressures $p_r$ and $p_t$ is what gives rise to the anisotropy of a fluid.  The difference between these two pressures is known as anisotropic factor $\sigma=p_t-p_r$. All pressures in question are functions of the radial coordinate. Thus the spherical symmetry is preserved, since we do not have inside the star tangential forces perpendicular to the radial direction.

The first study of the influence of anisotropic compact objects was analyzed by Bowers and Liang \cite{bowers_liang1974} in $1974$. In that work, the importance of the anisotropy of incompressible stars (stars with constant energy density, known also as Schwarzschild star) is studied through the generalization of the hydrostatic equilibrium equation, modified from it's original form to include the anisotropy effects. Using the anisotropic factor $\sigma=\alpha\,(p_r+\rho)\,(\rho+3\,p_r)\,e^{\lambda}\,r^n$, $\alpha$ and $n$ being constants (with $n>1$) and $e^{\lambda}$ a metric function, they found that the anisotropy has considerable effects on the maximum mass and on the surface redshift. As argued by the authors, the total radius of an incompressible star will be determined by the condition $p_r=0$, even though the anisotropic factor is nonnull on the surface, $\sigma_s\neq0$. In other words, an equilibrium solution can be found even with $\sigma_s\neq0$. In this work the radial stability is not investigated.

Shortly later, the effect of anisotropy on several macroscopic properties of neutron stars was analyzed by Hillebrandt in collaboration with Heintzmann in \cite{hillebrandt1975} and together with Steinmetz in \cite{hillebrandt1976}, considering that the anisotropic factor follows the function $\sigma= \beta\,p_r$, $\beta$ being a constant. Heintzmann and Hillebrandt in \cite{hillebrandt1975} found that for an arbitrarily large anisotropy there is no limiting mass and nor limiting redshift for neutron stars. After generalizing the Chandrasekhar radial pulsation equation for an object composed of an anisotropic fluid, Hillebrandt and Steinmetz in \cite{hillebrandt1976} analyzed the stability of the neutron stars. They show that a criterion in the stability against radial pulsation exists, similar to the one determined for isotropic models. In recently years, the influence of anisotropy both in nonradial oscillations, as in slowly rotating neutron stars, has been investigated considering the anisotropic factors $\sigma = \beta\,p_r(1-e^{ \lambda})$ in \cite{doneva_yazadjiev}, and with the one used in \cite{bowers_liang1974} (with $n=2$), in Ref. \cite{silva_macedo_berti_crispino2015}. Doneva and Yazadjiev in \cite{doneva_yazadjiev} analyze the influence of the anisotropy in the oscillation spectrum of a neutron star in the Cowling approximation. In that work, the authors found that the anisotropy could play an important role in the oscillation spectrum of a neutron star. Silva {\it et al.} in \cite{silva_macedo_berti_crispino2015} investigate the effects of the anisotropy on slowly rotating stars in General Relativity and in scalar-tensor theory. Silva and collaborators found that the anisotropy affects some physical properties of the neutron stars, such as the moment of inertia (in both theories) and the scalarization (a phase transition similar to spontaneous magnetization in ferromagnetic materials). The authors determined that the effects of scalarization grow (shrink) when the anisotropy increases (decreases).  In an anisotropic neutron star, unlike an anisotropic incompressible star, both the radial pressure and the energy density are zero on its surface, requiring that the anisotropy, in any of the forms previously given, is null on the star's surface $\sigma_s=0$.

From the afore cited works, we see that two possible forms of anisotropic factors can be analyzed, those that do not vanish on the star's surface $\sigma_s\neq0$ and those that do $\sigma_s=0$. In this work, we study for the first time the stability of anisotropic strange stars in these two cases. As is known, if  strange matter is the true ground state \cite{witten1984}, the compact objects known as pulsars would be strange stars rather than neutron stars \cite{itoh1970, Negreiro2009, Weber2007, Malheiro2003}. This leads to the need of studying anisotropic effects also in these types of compact objects. Possible sources of anisotropic pressure in quark matter could be color supercoductivity and viscosity, in the same way as superfluid and viscosity are in neutron stars, as has been discussed before.

We assume that the radial pressure and fluid energy density are connected by the MIT bag model equation of state. This equation of state was used in previous works, for instance, to study the radial oscillations of strange stars \cite{vath_chanmugam1992,benvenuto_horvath1991,gondek1999} and the stability of thin shell interfaces inside compact stars \cite{pereira_coelho_rueda2014} (review also \cite{pereira_rueda2015}). In turn, the anisotropic equation of state will be described by two possibles cases, one that follows the function $\sigma=\alpha\,(p_r+\rho)\,(\rho+3\,p_r)\,e^{\lambda}\,r^2$ and another one of the form $\sigma = \beta\,p_r(1-e^{ \lambda})$. Note that in a strange star, where the fluid is described by the MIT bag model equation of state, the energy density at the surface of the star is nonnull. This allows the first and second anisotropy factor to be respectively nonzero and zero at the star's surface, i.e., $\sigma_s\neq0$ and $\sigma_s=0$, respectively.

The article is organized as follows. In Sec.~\ref{gen_re_eq} we present the stress-energy tensor, both the stellar structure equations and the radial pulsation equations and their boundary conditions. In Sec.~\ref{eos_anis_carga} we make a brief description of the equation of state and of the two anisotropic factors considered. In Sec.~\ref{results} we present the numerical method used to integrate the stellar structure equations and the radial oscillation equations. We also present the scaling solution and results about the influence of the anisotropy in equilibrium and on the stability of strange quark stars for the two cases considered. Finally, in Sec.~\ref{conclusion} we conclude. It is worth mentioning that in this work the units $c=1=G$, where $c$ is the speed of light and $G$ is the gravitational constant, are considered. The metric is of signature $+2$ and Greek indices run from $0$ to $3$.

\section{General relativistic equations}\label{gen_re_eq}

\subsection{The stress-energy tensor and the background spacetime}

The matter contained in the spherical objects is described by the anisotropic fluid energy-momentum tensor, which, in the present work, it is written in the form \cite{herrera_barreto2013}
\begin{equation}\label{TEM}
T^{\mu}_\varphi=(\rho_{o}+p_{to})u^{\mu}u_{\varphi}+p_{to}g^{\mu}_{\varphi}+(p_{ro}-p_{to})k^{\mu}k_{\varphi},
\end{equation}
where the variables $\rho_o$, $p_{to}$, and $p_{ro}$ represent the energy density, the tangential pressure and the radial pressure, respectively. $u_\varphi$, $k_\varphi$, $g^{\mu}_\varphi$ stand respectively the fluid's four velocity, the radial unit vector and the metric tensor. As is known, the four velocity and the radial unit vector follow the conditions:
\begin{eqnarray}\label{vel_unit_condition}
 u_\varphi u^\varphi=-1,\;\;\;k_\varphi k^\varphi=1,\;\;\;{\rm and}\;\;\;k_\varphi u^\varphi=0.
\end{eqnarray}

The line element, which describes the interior of the object made of by the anisotropic fluid, is given by
\begin{equation}\label{metric}
ds^2=-e^{\nu_o}dt^2+e^{\lambda_o}dr^2+r^2d\theta^2+r^2\sin^2\theta d\phi^2,
\end{equation}
where the coordinates $(t,r,\theta,\phi)$ are the Schwarzschild coordinates.

It is worthwhile to mention that the variables of the metric $\nu_o$ and $\lambda_o$, and of the fluid $\rho_o$, $p_{ro}$, $p_{to}$ and etc. depend on the coordinates $t$ and $r$ only. When small radial perturbations in a given equilibrium configuration is studied, the variables that describe spacetime and the fluid are perturbed. Following the Chandrasekhar method, these variables are decomposed in the form
\begin{equation}\label{perturbation}
g_o(t,r)=g(r)+\delta g(t,r),
\end{equation}
where $g(r)$ denotes the unperturbed potential metrics and physical quantities that depend only on the variable $r$, i.e., $g(r)$ represents the variables in a static equilibrium configuration. In turn $\delta g(t,r)$ represents the Eulerian perturbations that depend on the variables $t$ and $r$.

\subsection{Stellar structure equations}

The system of equations used to study the spherically symmetric configurations of an anisotropic fluid in a static regime, namely, in an unperturbed regime ($\delta g=0$), are given by the following relations:
\begin{eqnarray}
&&\frac{dm}{dr}=4\pi r^2\rho,\label{mo}\\
&&\frac{dp_r}{dr}=-(p_r+\rho)\left(4\pi r p_r+\frac{m}{r^2}\right)e^{\lambda}+\frac{2\sigma}{r},\label{tov}\\
&&\frac{d\nu}{dr}=-\frac{2}{(p_r+\rho)}\frac{dp_r}{dr}+\frac{4\sigma}{r(p_r+\rho)},\label{nuo}
\end{eqnarray}
where the metric component $e^{\lambda}$ is given by the equality:
\begin{equation}\label{lambdao}
e^{\lambda}=\left(1-\frac{2m}{r}\right)^{-1}.
\end{equation}
As usual, $m$ denotes the mass within the sphere radius $r$. Equation \eqref{tov} represents the hydrostatic equilibrium equation modified to the inclusion of the anisotropy factor $\sigma=p_t-p_r$ \cite{bowers_liang1974}. This TOV equation matches with the hydrostatic equilibrium equation for an isotropic fluid \cite{tolman,oppievolkoff} by taking $\sigma=0$.   

\subsection{Radial pulsation equations}

With the purpose of determining the Eulerian perturbations, firstly, the nonzero components of four-velocity and of the radial unit vectors are defined. The components considered have a similar form to those presented in \cite{horvat_ilijic_marunovic}, in the study of radial oscillations of anisotropic polytropic stars. It is worth noting that the definitions considered for the nonzero components of four-velocity and of the radial unit vectors satisfy the conditions shown in \eqref{vel_unit_condition}. After, we decompose the spacetime and the fluid variables of the form presented in \eqref{perturbation}. The definitions and the decompositions aforesaid are inserted in the field equations, while by maintaining only the first-order terms, we find the perturbations:
\begin{eqnarray}
&&\delta\lambda=-8\pi(p_r+\rho)re^{\lambda}\zeta,\label{lambda_per}\\
&&\frac{e^{-\lambda}}{2r}\frac{\partial(\delta\nu)}{\partial r}=4\pi\left(\delta p_r-(p_r+\rho)\zeta\left(\frac{d\nu}{dr}+\frac{1}{r}\right)\right),\label{nu_per}\\
&&\delta\rho=-\zeta\frac{d\rho}{dr}-\frac{(p_r+\rho)e^{\frac{\nu}{2}}}{r^2}\frac{d}{dr}\left[\frac{r^2\zeta}{e^{\frac{\nu}{2}}}\right]-\frac{2\sigma\zeta}{r},\label{rho_per}\\
&&\delta p_r=-\zeta\frac{dp_r}{dr}-\frac{\Gamma p_r e^{\frac{\nu}{2}}}{r^2}\frac{d}{dr}\left[\frac{r^2\zeta}{e^{\frac{\nu}{2}}}\right]-\frac{2\Gamma p_r\zeta}{r}\frac{\sigma}{(p_r+\rho)},\label{p_per}
\end{eqnarray}
with $\Gamma$ being the adiabatic index, defined as $\Gamma=\left(1+\frac{\rho}{p_r}\right)\frac{dp_r}{d\rho}$, and with $\zeta$ being the ``Lagrangian displacement'' with respect to the worldtime $t$ defined by $v=\frac{\partial\zeta}{\partial t}$. The equation which determines the radial pulsation is found using  the linearized form of energy-momentum tensor conservation, the previous results for the perturbed quantities, and taking into account that all perturbations have a time dependence of the form $e^{i\omega t}$, where $\omega$ is known as  eigenfrequency or oscillation frequency. Thus, we have:
\begin{eqnarray}\label{RO}
&&\omega^2(p_r+\rho)e^{\lambda-\nu}\zeta-8\pi(p_r+\rho)\zeta e^{\lambda}\left(\sigma+p_r\right)+\frac{8\sigma\zeta}{r^2}+\frac{\zeta(p_r+\rho)}{4}\left(\frac{d\nu}{dr}\right)^2+\frac{2\delta\sigma}{r}-\frac{4\zeta}{r}\frac{dp_r}{dr}\nonumber\\
&&+e^{-\frac{\lambda}{2}-\nu}\frac{d}{dr}\left(e^{\frac{\lambda}{2}+\nu}\left(\frac{\Gamma p_r e^{\frac{\nu}{2}}}{r^2}\frac{d}{dr}\left(\frac{r^2\zeta}{e^{\frac{\nu}{2}}}\right)+\frac{2\zeta\sigma}{r}\left(\frac{\Gamma p_r}{p_r+\rho}+1\right)\right)\right)=0.
\end{eqnarray}
This is the radial pulsation equation, i.e., the Chandrasekhar pulsation equation,  modified to the inclusion of the anisotropy factor \cite{Dev2003,dev_gleiser2004}. It is important to mention that the radial pulsation equation for an isotropic fluid \cite{Chandrasekhar1964-a,Chandrasekhar1964-b} can be recovered taking into account $\sigma= 0$ in equation \eqref{RO}. It is widely known that equation \eqref{RO} can be placed in a  form that allows a more suitable numerical integration. It can be derived in a system of two first-order equations, of the form 
\begin{eqnarray}
&&\frac{d\xi}{dr}=\frac{\xi}{2}\frac{d\nu}{dr}-\frac{1}{r}\left[3\xi+\frac{\Delta p_r}{\Gamma p_r}+\frac{2\xi\sigma}{\Gamma p_r}\left[\frac{\Gamma p_r}{p_r+\rho}+1\right]\right],\label{ro1}\\
&&\frac{d(\Delta p_r)}{dr}=e^{\lambda-\nu}(p_r+\rho)\omega^2 \xi r -4\xi \frac{dp_r}{dr}+\frac{8\sigma\xi}{r}-\left(\frac{1}{2}\frac{d\nu}{dr}+4\pi re^{\lambda}(p_r+\rho)\right)\Delta p_r\nonumber\\
&&+\frac{\xi r(p_r+\rho)}{4}\left(\frac{d\nu}{dr}\right)^2-8\pi(p_r+\rho)r\xi e^{\lambda}\left(\sigma+p_r\right)+\frac{2\delta\sigma}{r},\label{ro2}
\end{eqnarray}
with $\xi=\zeta/r$ and $\Delta p_r$ representing respectively the relative radial displacement and the Lagrangian perturbation. The set of equations \eqref{ro1} and \eqref{ro2} can be reduced to the first-order equations presented in \cite{gondek1997,lugones2010} taking $\sigma=0$. 

\subsection{About the boundary conditions}

\subsubsection{Stellar structure equations}

Note that the set of equations \eqref{mo}, \eqref{tov}, \eqref{nuo} and \eqref{lambdao} are not sufficient to solve for the six unknown functions $\lambda$, $\nu$, $m$, $p_r$, $\rho$,  and $\sigma$. To complete the system of equations, it is usually considered an equation of state relating the energy density with the radial pressure of the fluid and a relation defining the anisotropy. In our two forms for $\sigma(r)$, knowing $p_r$ and $\rho$, $\sigma$ is known 

Once completed the system of equations, some boundary conditions are required to solve the equations \eqref{mo}-\eqref{lambdao} along the radial coordinate $r$, from the center toward the star's surface. In the center $r=0$ of the object is considered: 
\begin{eqnarray}\label{boundary1}
m(0)=0,\hspace{+0.5cm}\rho(0)=\rho_{c},\hspace{+0.5cm}p_r(0)=p_{rc},\hspace{+0.5cm}\sigma(0)=0,\hspace{+0.5cm}\lambda(0)=0,\hspace{+0.5cm}\nu(0)=\nu_c,
\end{eqnarray}
and the surface of the object $r=R$ is found when
\begin{equation}\label{boundary2}
p_r(R)=0.
\end{equation}
At this point the interior solution will match smoothly with the Schwarzschild metric outside the star, indicating that the potential metrics of the interior and the exterior metric are related as
\begin{eqnarray}\label{condition}
e^{\nu(R)}=\frac{1}{e^{\lambda(R)}}=1-\frac{2M}{R},
\end{eqnarray}
being $M$ the total mass of the star. The relation \eqref{condition} gives the boundary condition for the functions $\nu$ and $\lambda$ at the star surface.

\subsubsection{Radial pulsation equations}

To integrate equations \eqref{ro1} and \eqref{ro2}, from the center to the surface of the star, some boundary conditions must be defined. In order to have a regular solution in the center of the star, it is required that for $r\rightarrow0$ the coefficient in Eq.~\eqref{ro1} must vanish:
\begin{equation}\label{bc_oscillation}
(\Delta p_r)_{\rm center}=\left[-2\xi\sigma\left[\frac{\Gamma p_r}{p_r+\rho}+1\right]-3\xi\Gamma p_r\right]_{\rm center}.
\end{equation}
In the center of the star, for normalized eigenfunctions, we have $\xi(r=0)=1$. The null radial pressure on the surface of the stars $p_r(R)=0$ implies in
\begin{equation}\label{bc_oscillation1}
(\Delta p_r)_{\rm surface}=0.
\end{equation}

\section{Equation of state and anisotropic profile}\label{eos_anis_carga}

\subsection{Equation of state}

To describe strange quark matter, the MIT bag model is considered. It is the simplest equation of state used to study the equilibrium configuration of stellar objects whose matter content consists only of up, down and strange quarks. This equation of state takes into account that these quarks are massless and non-interacting quarks confined by a bag constant ${\cal B}$. For the anisotropic fluid studied, we consider that the energy density and the radial pressure of the fluid are connected through the relation:
\begin{equation}\label{eos} 
\rho=3\,p_r+4{\cal B},
\end{equation} 
being ${\cal B}$ the bag constant. The theoretical possibility that the strange quark matter might be the true ground state of strongly interacting matter has been proposed by Witten \cite{witten1984}. If this conjecture is correct, the strange matter would manifest in compact stars. Farhi and Jaffe showed \cite{farhi_jaffe1984} that, for massless and non-interacting quarks, the Witten conjecture is verified for a bag constant approximately between the values $57\,{\rm MeV/fm^3}$ and $94\,{\rm MeV/fm^3}$. For the study of strange stars, we consider ${\cal B}=60\,{\rm MeV/fm^3}$.

\subsection{Anisotropic profile}

We consider two anisotropic factors depending on local variables, i.e., depending on some quantities that refer to the state of the fluid and others that can be derived from the geometry at the particular point of the spacetime. Then, we use the anisotropic profiles
\begin{eqnarray}
&&\sigma=\alpha(\rho+p_r)(\rho+3\,p_r)r^2 e^{\lambda},\label{eos_anis2}\\
&&\sigma=\beta p_r (1-e^{-\lambda}),\label{eos_anis}
\end{eqnarray}
being $\alpha$ and $\beta$ the anisotropic constants. Henceforward, the results found using the anisotropic factor (\ref{eos_anis2}) (AF $1$) will be named case $1$, and those found with the anisotropic factor (\ref{eos_anis}) (AF $2$) as $2$. The AF $1$ was used by the first time in \cite{bowers_liang1974}. This anisotropic factor was employed later, for instance, to analyze the adiabatic contraction of an anisotropic sphere \cite{herrera1979} and to investigate the influence of the anisotropy in slowly rotating neutron stars \cite{silva_macedo_berti_crispino2015}. The AF $2$ was introduced in \cite{cattoen_faber_visser} and used later, for example, in \cite{horvat_ilijic_marunovic} to develop an analytical study of the radial oscillations of polytropic stars made by anisotropic fluid, in studying the nonradial oscillations of neutron stars \cite{doneva_yazadjiev}, the influence of the anisotropy on magnetic field structure \cite{folomeev2015}, and in analyzing the slowly rotating anisotropic neutron stars \cite{silva_macedo_berti_crispino2015}. 

In order to handle a single anisotropic constant, we consider $\alpha=\frac{\kappa}{1.7}$, in the case $1$, and $\beta=\kappa$, in the case $2$. The particular form of the factor $\frac{\kappa}{1.7}$ in the case $1$, it is chosen with the aim of finding very similar maximum stellar masses using the AF $1$ and AF $2$ for any value of $\kappa$. It is important to mention that the two anisotropy factors assures regularity of Eqs.~\eqref{tov} and \eqref{nuo} and of the first-order terms of perturbations \eqref{rho_per} , \eqref{p_per} and \eqref{RO} at the star center.

\section{Equilibrium and stability of anisotropic strange stars}\label{results}

\subsection{Numerical method}

The stellar structure equations and the radial pulsation equations with their boundary conditions are integrated from the center toward the surface of the star. To analyze the equilibrium configuration of a star, we integrate numerically the stellar structure equations using the Runge-Kutta 4th-order method for each $\rho_c$ and $\kappa$. After obtaining the coefficients of the radial pulsation equations for a given $\rho_c$ and $\kappa$, the radial oscillation equations are solved by means of the shooting method. This process begins considering a test value for $\omega^2$. If after the integration the condition \eqref{bc_oscillation1} is not satisfied, $\omega^2$ is corrected until matching this condition in the next integration. The values of $\omega^2$ that satisfy the condition $\Delta p=0$ are known as eigenfrequencies or frequency of oscillations.

\subsection{Scaling solution of stellar structure equations and of radial pulsation equations}

It has previously been observed that when the fluid contained in a star follows a linear relation between the energy density and the radial pressure, such as the MIT bag model equation of state $\rho=3p_r+4{\cal B}$, the equation of stellar structure and the radial pulsation equations admit a scaling law for various properties of the stars (see e.g. \cite{witten1984,glendenningbook}). I.e., if the properties of a star are known for a given value of ${\cal B}$, such quantities can be easily obtained to another value ${\cal B}'$. For anisotropic stellar objects, the law of scale is possible using the variables:
\begin{eqnarray}\label{scale}
{\bar p}_r=\frac{p_r}{\cal B};\hspace{0.3cm}{\bar \rho}=\frac{\rho}{\cal B};\hspace{0.3cm}{\bar\sigma}=\frac{\sigma}{\cal B};\hspace{0.3cm}{\bar m}=m\sqrt{\cal B};\hspace{0.3cm}{\bar r}=r\sqrt{\cal B};\hspace{0.3cm}{\bar\omega}=\frac{\omega}{\sqrt{\cal B}};\hspace{0.3cm}{\bar\xi}=\frac{\xi}{e};\hspace{0.3cm}{\bar\Delta p}=\frac{\Delta p}{f},
\end{eqnarray}
where $f={\cal B}e$, and with $f$ and $e$ being positive and nonzero. Using the scaling law shown in \eqref{scale}, both the stellar structure equations and the radial oscillations equations maintain its original form. Once known the properties of a star for a particular bag constant value (${\cal B}$), the properties of other different star with a different bag constant value (${\cal B}'$) can be found through the scale:
\begin{eqnarray}
\frac{\rho_c'}{{\cal B}'}=\frac{\rho_c}{{\cal B}};\,\,\,\,M'\sqrt{{\cal B}'}=M\sqrt{\cal B};\,\,\,\,R'\sqrt{{\cal B}'}=R\sqrt{\cal B};\,\,\,\,z'=z;\,\,\,\,\,\frac{\omega'}{\sqrt{{\cal B}'}}=\frac{\omega}{\sqrt{\cal B}},
\end{eqnarray}
where the constant $\rho_c$ represents the central energy density and $z$ the redshift.

\subsection{Equilibrium and stability of anisotropic strange stars with fixed $\kappa$}

Firstly, we will present the anisotropic effect in the energy and the radial pressure inside the strange stars. After that, its effects in the equilibrium configurations and stability are presented. The anisotropic effects are analyzed considering a fixed $\kappa$ .

\subsubsection{The influence of the anisotropy in the energy density and radial pressure of the fluid with fixed $\kappa$}

\begin{figure}[tbp]
\centering
\includegraphics[width=0.32\linewidth,angle=0]{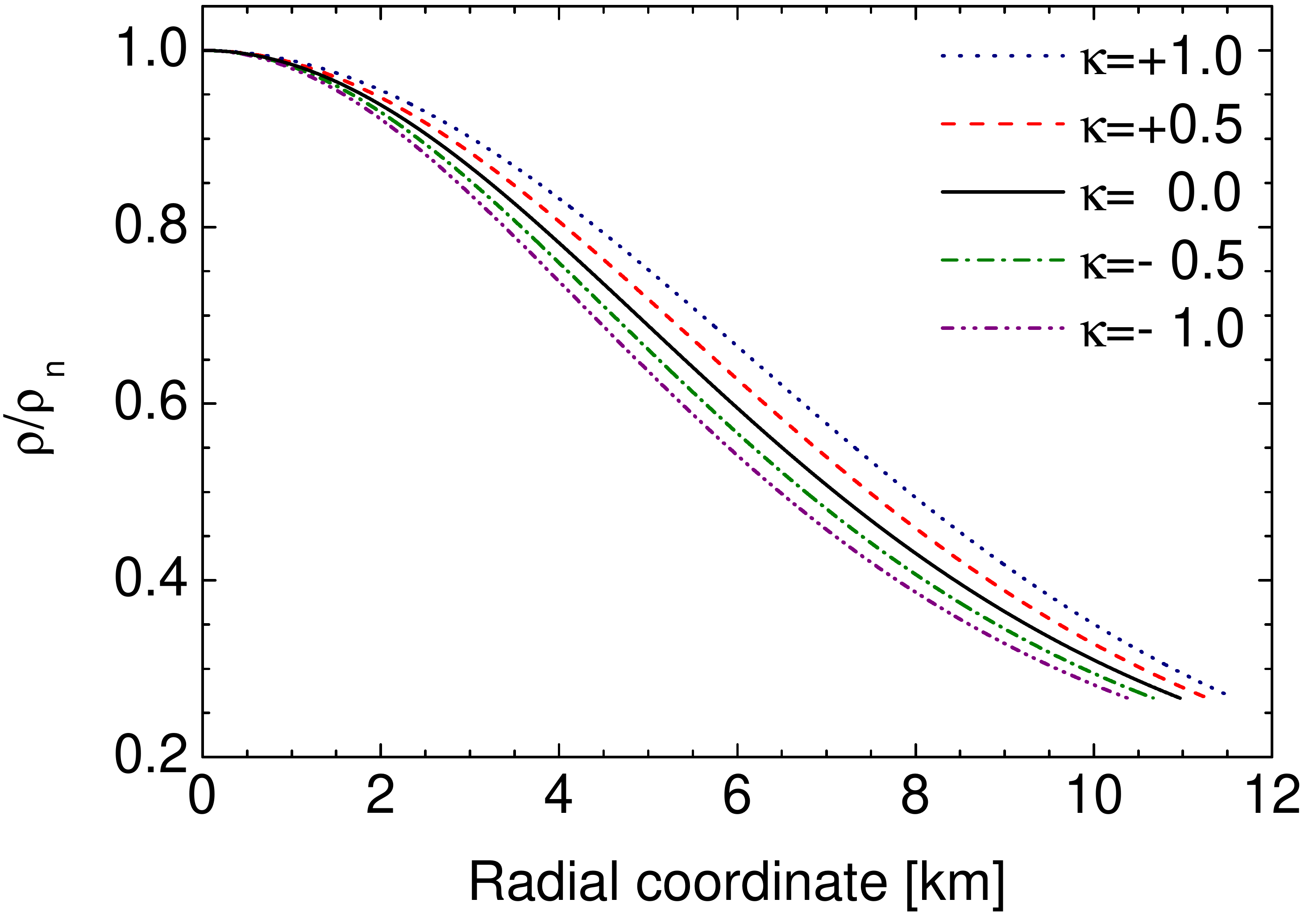}
\hfill
\includegraphics[width=0.32\linewidth,angle=0]{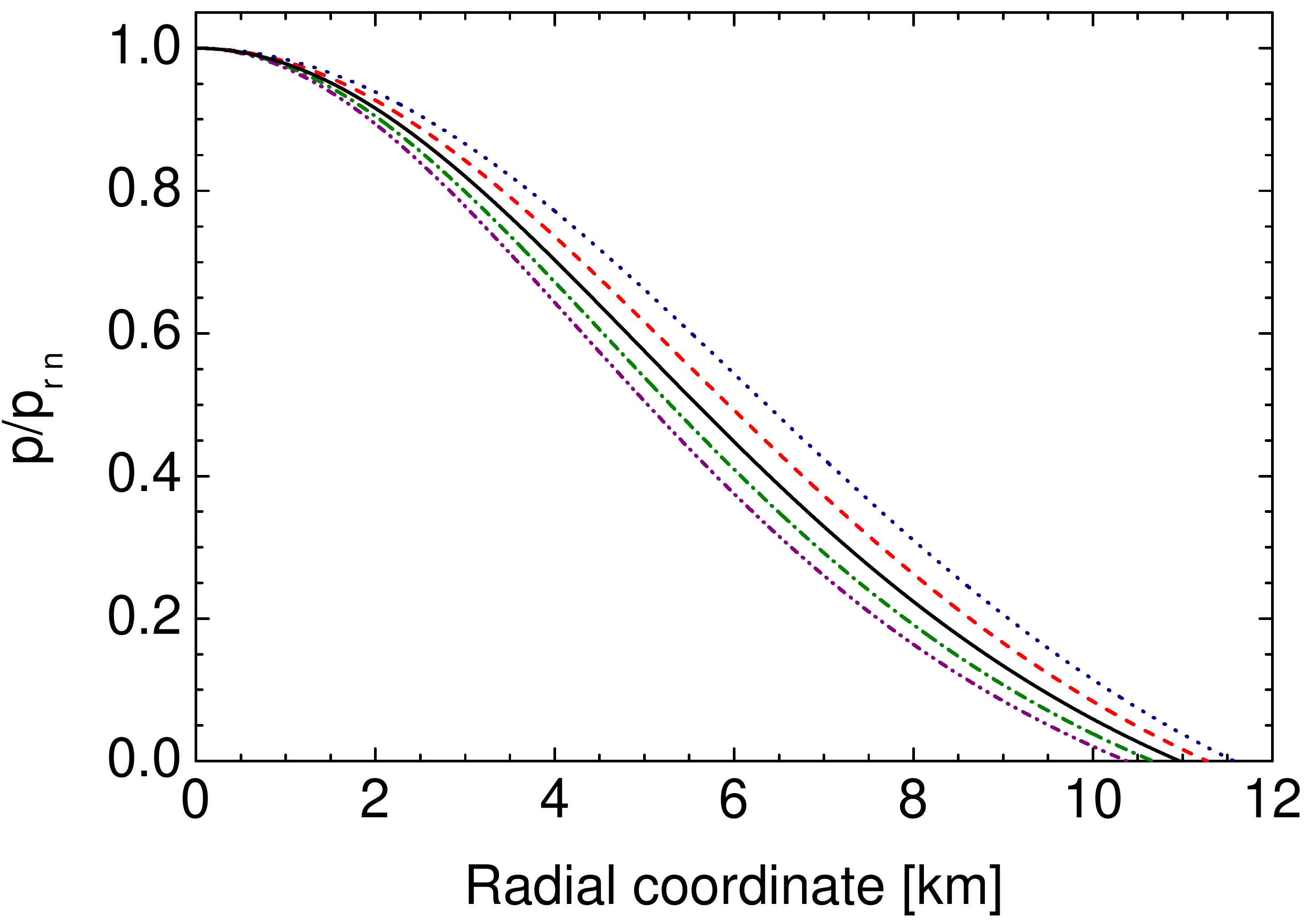}
\hfill
\includegraphics[width=0.32\linewidth,angle=0]{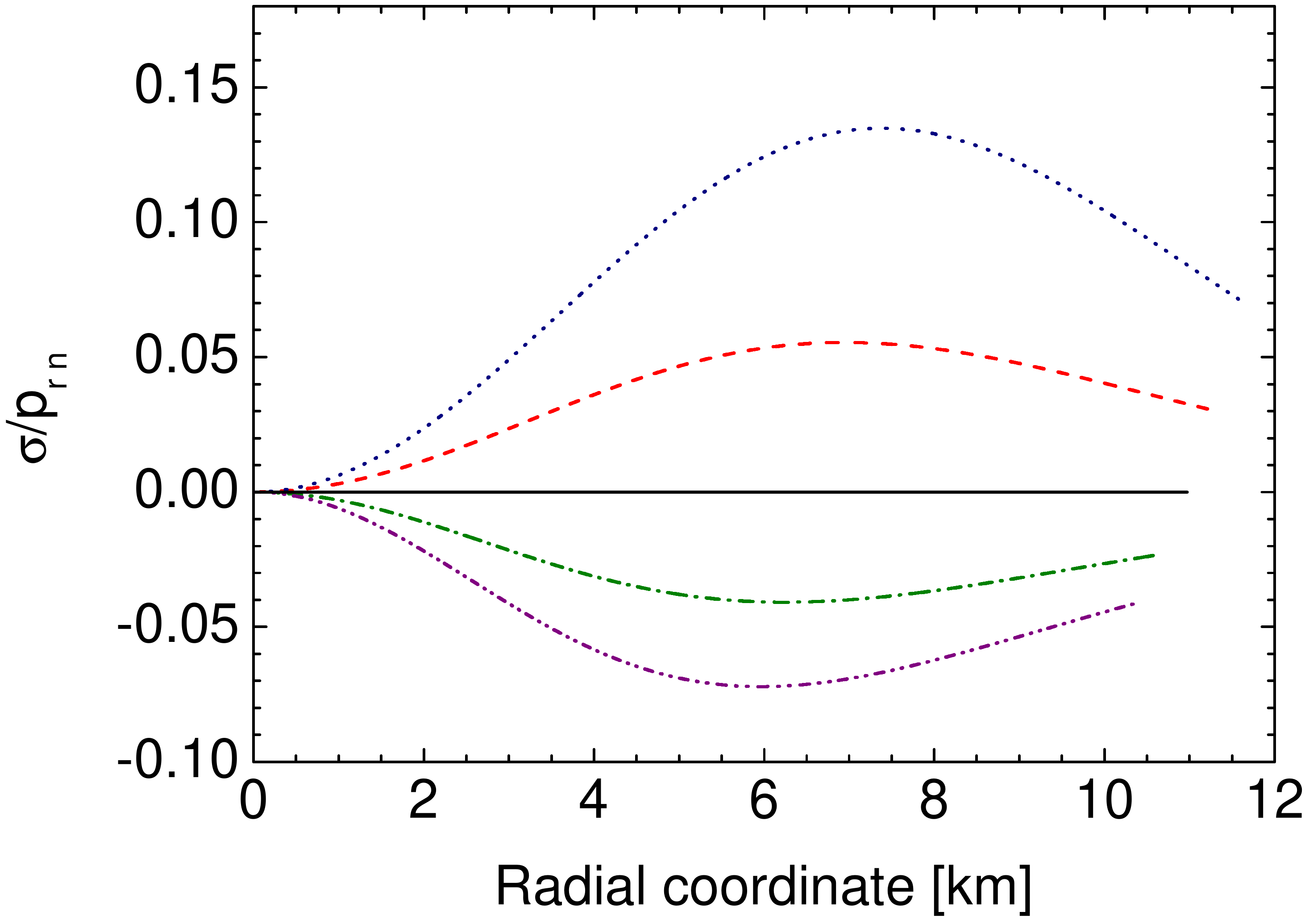}
\    
\includegraphics[width=0.32\linewidth,angle=0]{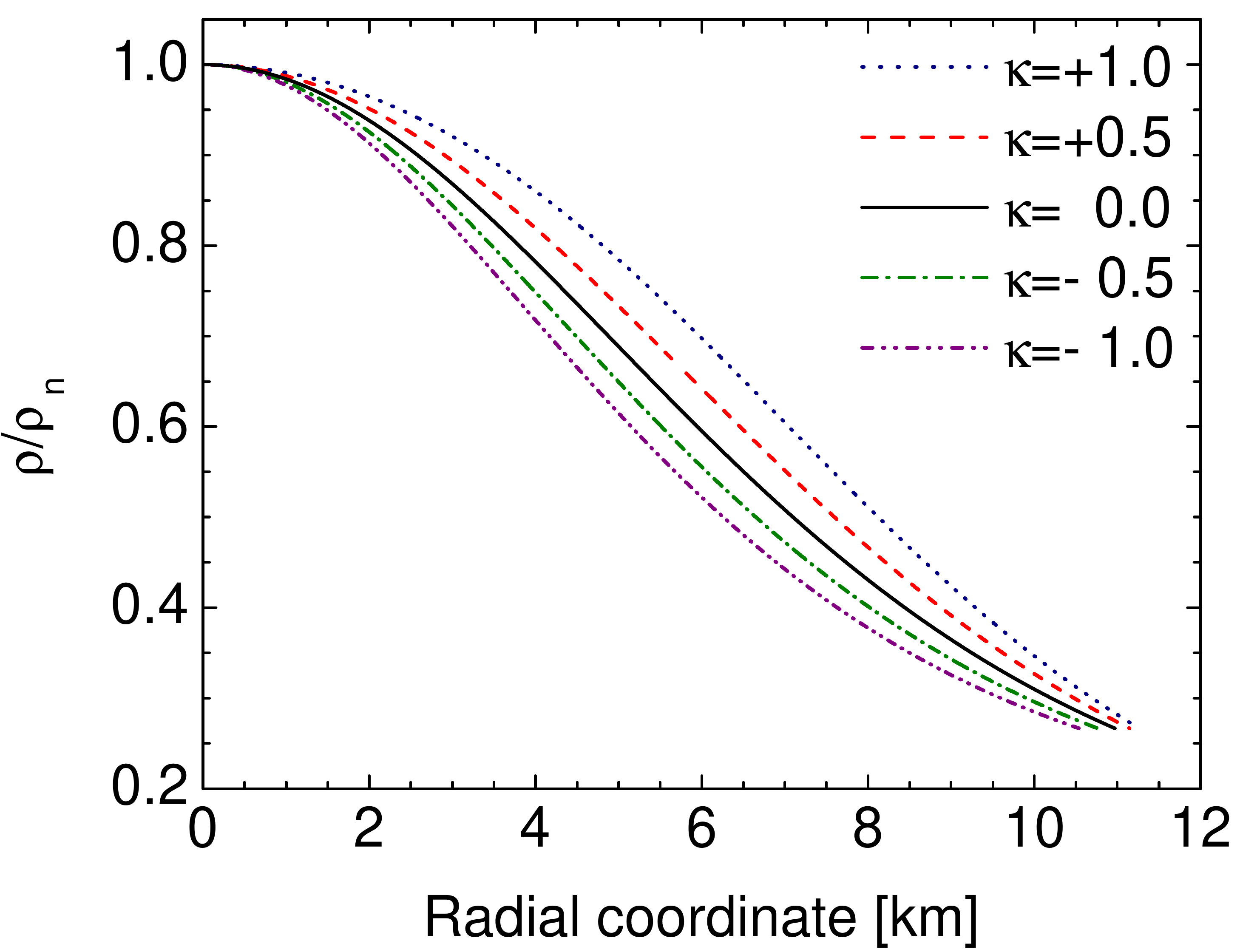}
\hfill
\includegraphics[width=0.32\linewidth,angle=0]{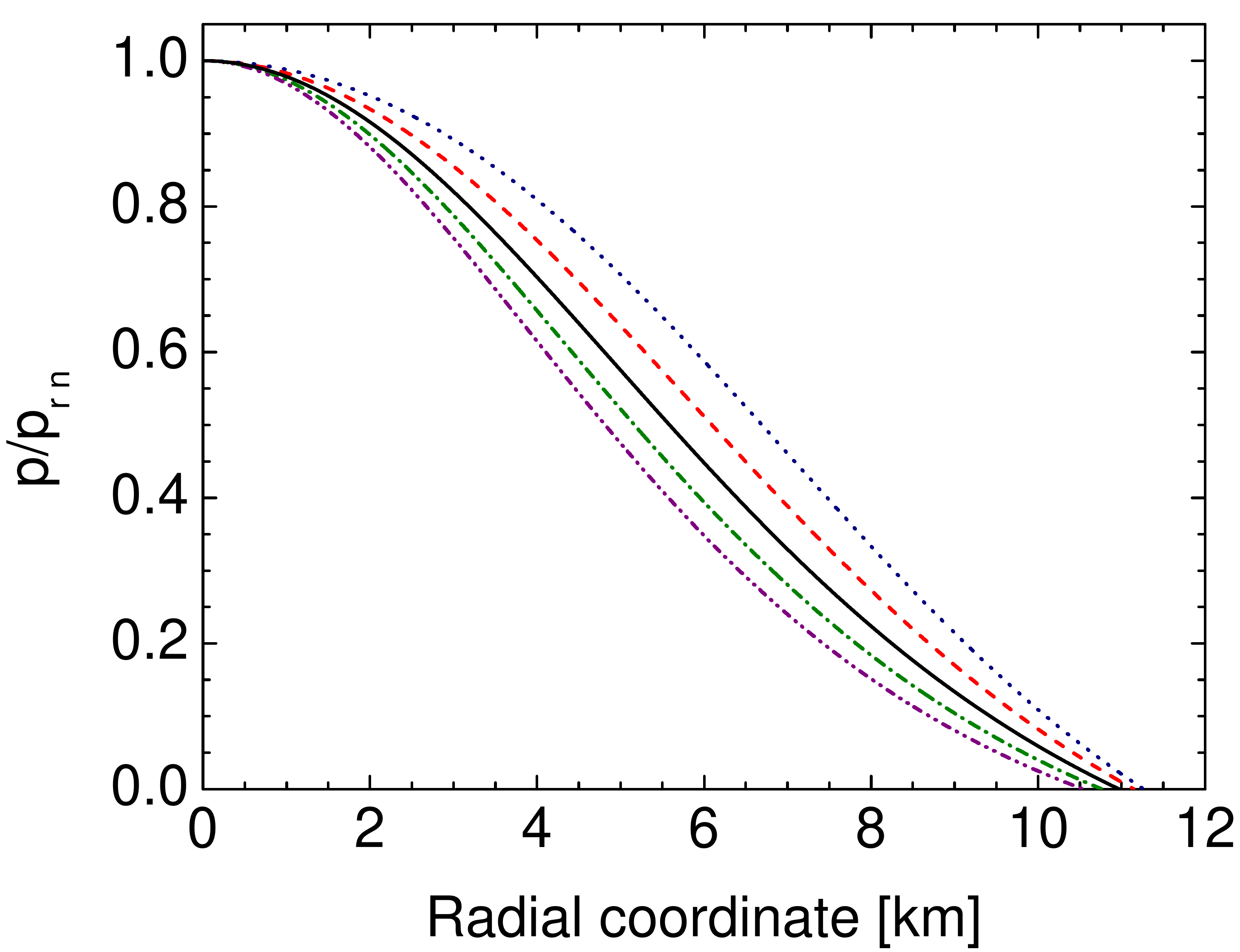}
\hfill
\includegraphics[width=0.32\linewidth,angle=0]{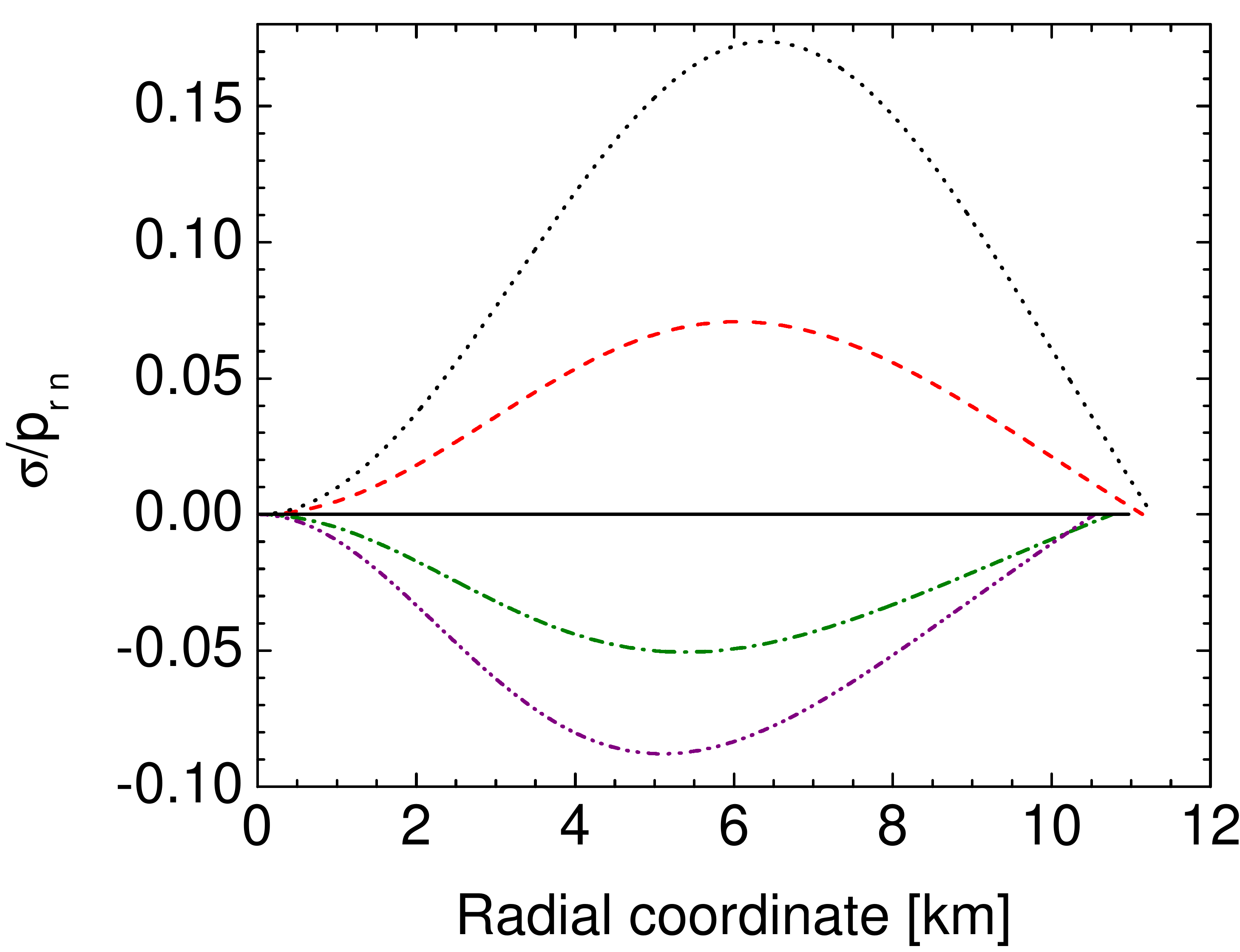}
\caption{\label{rho_p_r}The energy density, radial pressure and anisotropy factor in their normalized forms as a function of the radial coordinate are shown on the panels on the top  and on the bottom, where the anisotropic factors $1$ and $2$  are respectively used. $\rho_c=900\,{\rm MeV/fm^3}$ and $p_{rc}=220\,{\rm MeV/fm^3}$ are employed in the panels above and below. The energy density and radial pressure of normalization are respectively $\rho_n=900\,{\rm MeV/fm^3}$ and $p_{rn}=220\,{\rm MeV/fm^3}$.}
\end{figure}

The behavior of the energy density $\rho/\rho_n$, the radial pressure $p_{r}/p_{rn}$ and the anisotropy $\sigma/p_{rn}$ in the radial coordinate are presented in Fig.~\ref{rho_p_r} for five different values of $\kappa$. On the top panels and on the bottom panels are considered respectively the AF $1$ and AF $2$. The central energy density and the central radial pressure considered both on the panels above as the ones below are $\rho_c=900\,{\rm MeV/fm^3}$ and $p_{rc}=220\,{\rm MeV/fm^3}$. The normalization factor for the energy density is $\rho_n=900\,{\rm MeV/fm^3}$, and for the radial pressure and anisotropy factor is $p_{rn}=220\,{\rm MeV/fm^3}$. Note in Fig.~\ref{rho_p_r} that in the two cases under analyze, both the energy density and the radial pressure profile in a star are affected due to the anisotropy. The energy density and the radial pressure in the interior of a star increases (decreases) with the increment (diminution) of $\kappa$. On the other hand, in the case $1$, we can observe that the anisotropy at the star's surface does not vanish, $\sigma_s\neq0$, in turn, in the case $2$, we note that the anisotropy vanishes at this point, $\sigma_s=0$. 

\subsubsection{Equilibrium of anisotropic strange stars with fixed $\kappa$}

\begin{figure}[tbp]
\centering
\includegraphics[width=0.45\linewidth,angle=0]{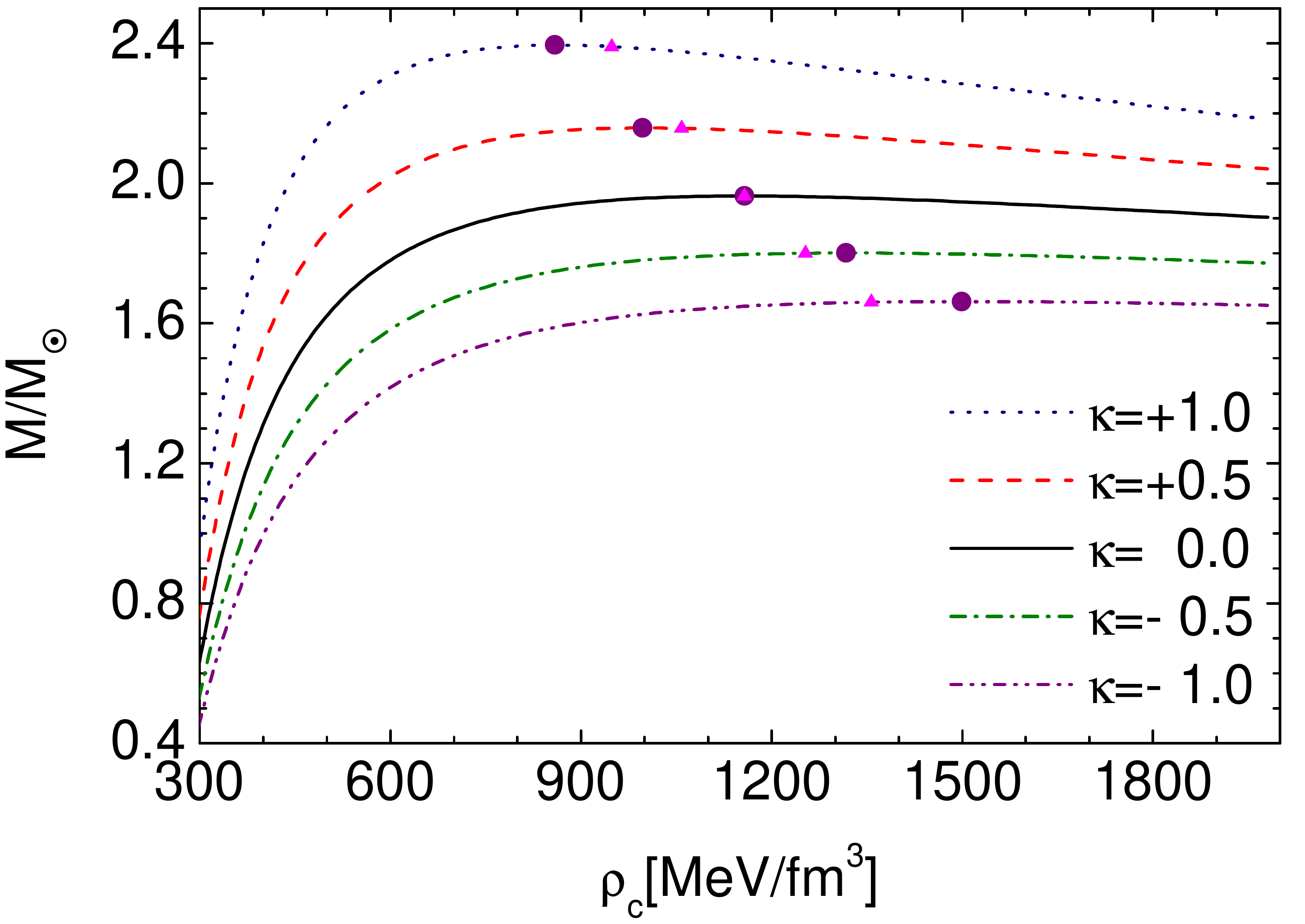}
\hfill
\includegraphics[width=0.45\linewidth,angle=0]{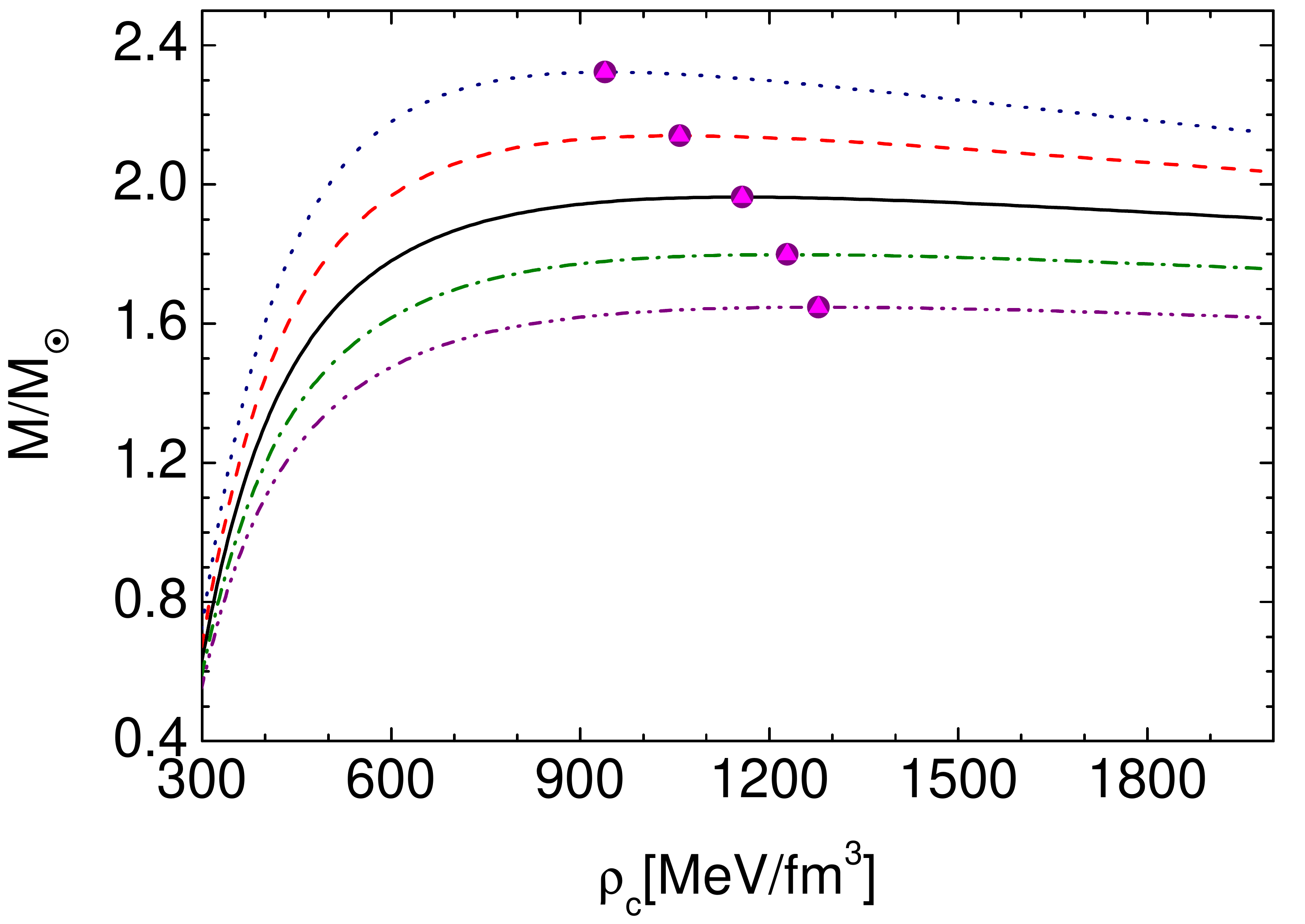}
\caption{\label{M_rhoc_Carga0}The total mass $M/M_{\odot}$ as a function of the central energy density ${\rho_c}$ for five different values of $\kappa$. The AF $1$ and AF $2$ are used respectively on the left panel and on the right panel. The full circles and the full triangles represent respectively the maximum mass points and the zero oscillation frequencies.}
\end{figure}

The changing of the mass $M/M_{\odot}$ against the central energy density $\rho_c$ is shown in Fig.~\ref{M_rhoc_Carga0} for five different values of $\kappa$, considering the AF $1$ on the left panel and the AF $2$ on the right one. In Fig.~\ref{M_rhoc_Carga0}, the values of the central energy density considered are between the values $300$ and $2000\,{\rm MeV/fm^3}$. The spheres and the triangles on the curves represent the points where the maximum mass points $M_{\rm max}/M_{\odot}$ and the zero frequency of oscillations $\omega=0$ are found. In the case $1$, where $\sigma_s\neq0$, we can note that $M_{\rm max}/M_{\odot}$ and $\omega=0$ are obtained at different central energy densities. From this we understand that the condition to distinguish a stable (unstable) equilibrium region in a sequence of equilibrium configurations $dM/d\rho_c>0$ ($dM/d\rho_c<0$), it is a necessary condition but not sufficient. In turn, in the case $2$, where $\sigma_s=0$, we can clearly observed in each curve that the maximum mass point and the zero frequency of oscillation are found using the same central energy density $\rho_c^{*}$. In this last case, the equilibrium configurations lying in the region where $\rho_c<\rho_c^{*}$ are stable, due to the fact that for those central energy densities, $\omega^2>0$. Thus, in case 2 we can conclude that the regions composed by stable and unstable equilibrium configurations can be always be recognized through the conditions $dM/d\rho_c>0$ and $dM/d\rho_c<0$, respectively. A similar result was found in \cite{hillebrandt1976}, studying the effect of the anisotropy $\sigma= \beta\,p_r$ in the stability of neutron stars. Due to the anisotropic factor considered in \cite{hillebrandt1976}, where the anisotropy at the star's surface is null $\sigma_s=0$.

\begin{figure}[tbp]
\centering
\includegraphics[width=0.45\linewidth,angle=0]{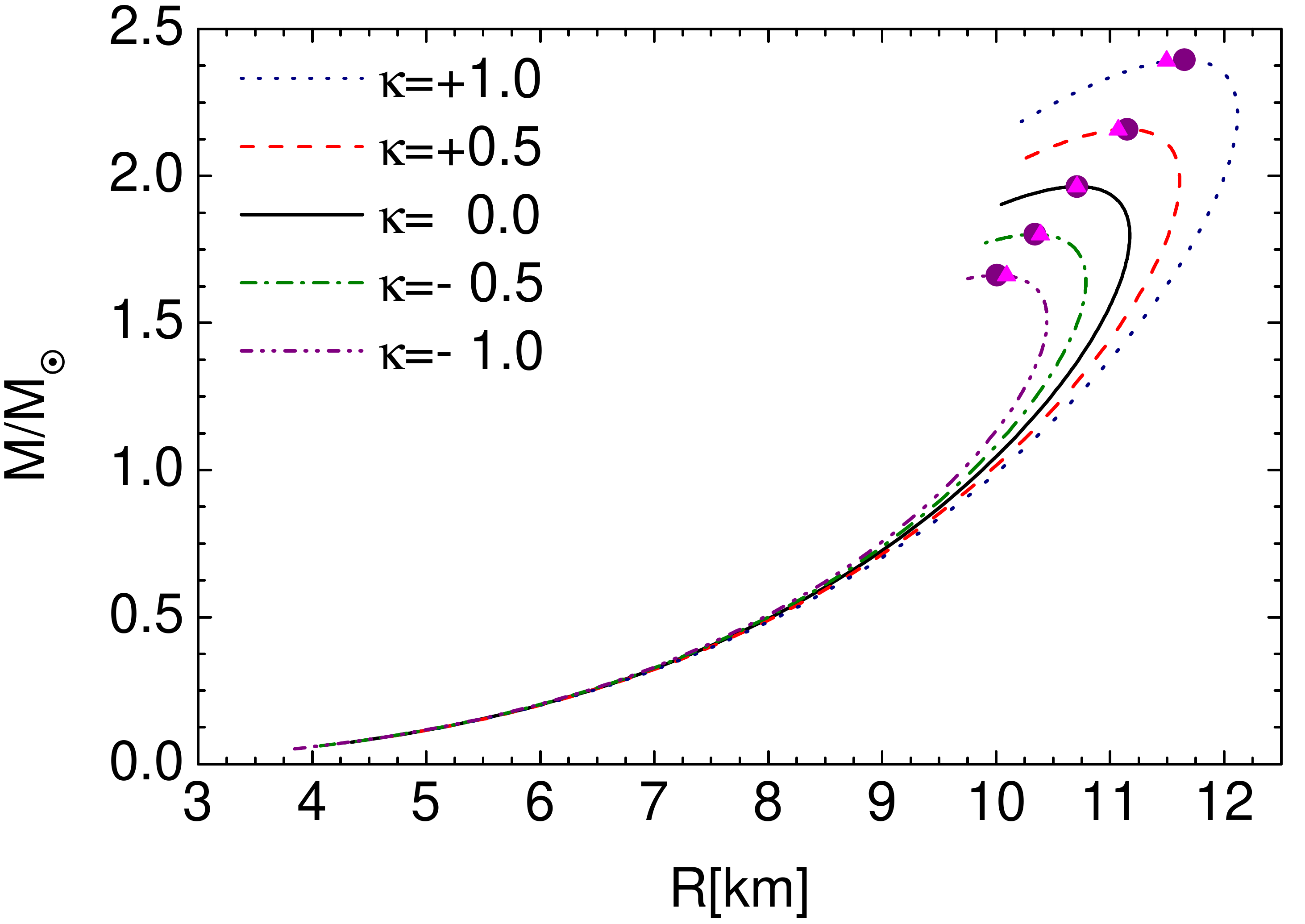}
\hfill
\includegraphics[width=0.45\linewidth,angle=0]{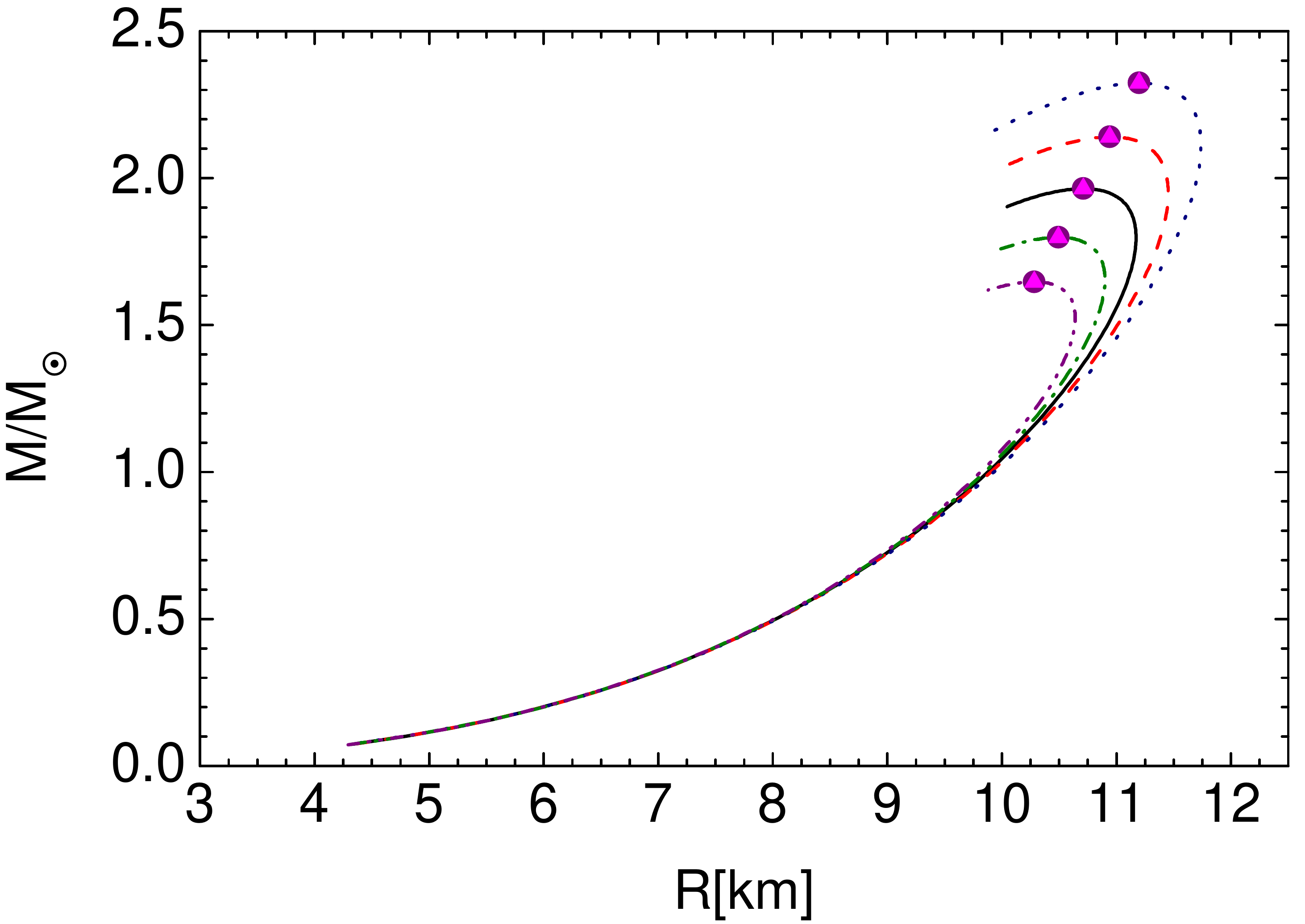}
\caption{\label{M_R_Carga0}The mass as a function of the radius for different values of $\kappa$. On the left panel and on the right panel are used respectively AF $1$ and AF $2$. The maximum mass points and the zero oscillation frequencies values are indicated by full circles and full triangles, respectively.}
\end{figure} 

The behavior of the total mass, normalized in solar masses $M_{\odot}$, as a function of the total radius of the star $R$ for five different values of $\kappa$ is presented in Fig. \ref{M_R_Carga0}. On the left and on the right panels are considered respectively the anisotropic factors AF $1$ and AF $2$. From figures 2 and 3, we can observe that the anisotropy affects the equilibrium configuration of the star. Analyzing the maximum mass and its respective total radius, we found that these values change in a considerable percentage with $\kappa$. Once $\kappa>0$, for a larger $\kappa$ the star needs to have larger maximum mass and total radius to attain its equilibrium configuration. The increment of the mass with $\kappa$ can be explained analyzing the increment of $\sigma$ with $\kappa$. A larger anisotropy $\sigma$ helps the fluid pressure to support stars with larger mass and radius against the collapse. In turn, a lower $\sigma$ produce equilibrium configuration with lower mass and radii. 

\begin{figure}[tbp]
\centering
\includegraphics[width=0.45\linewidth,angle=0]{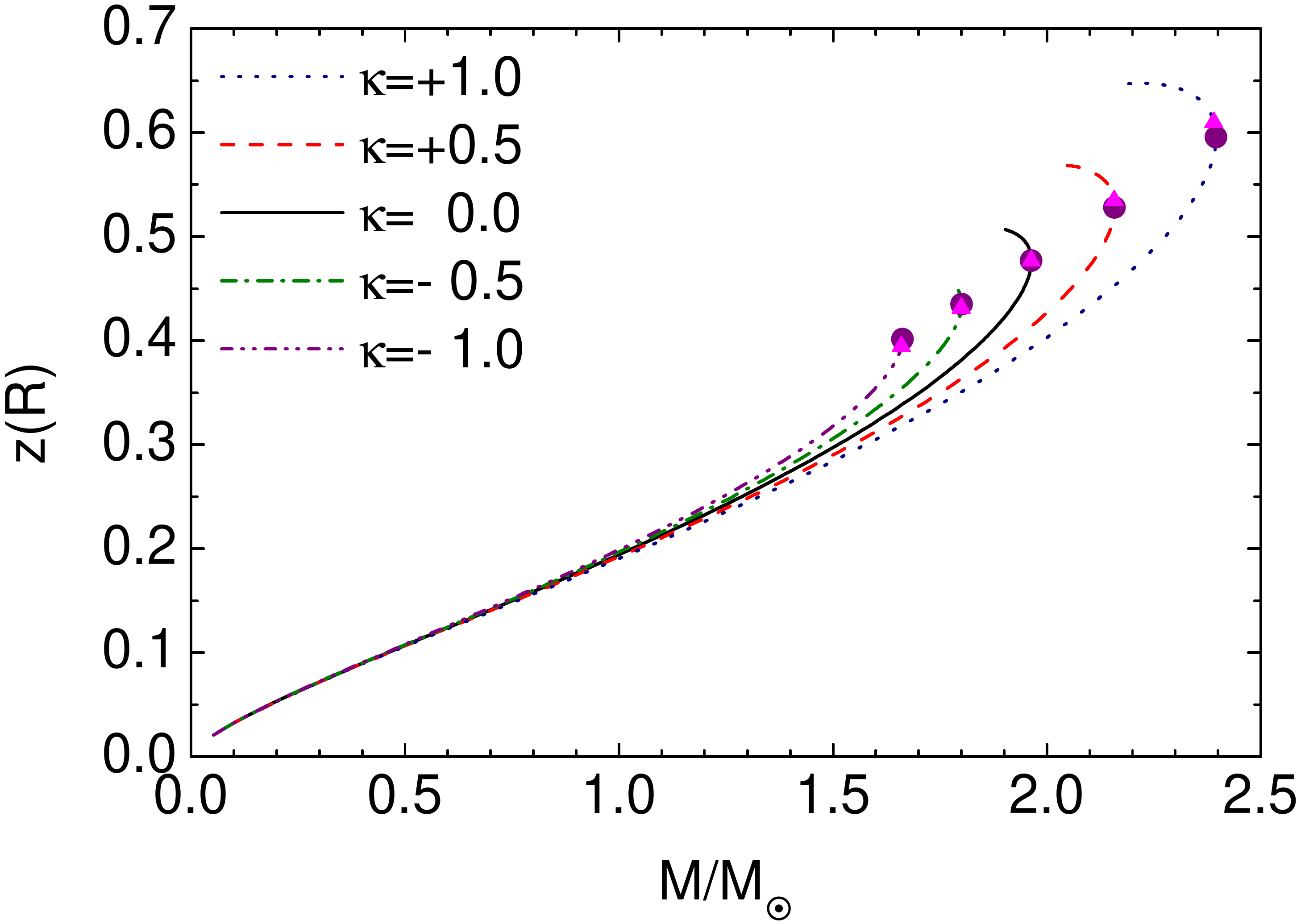}
\hfill
\includegraphics[width=0.45\linewidth,angle=0]{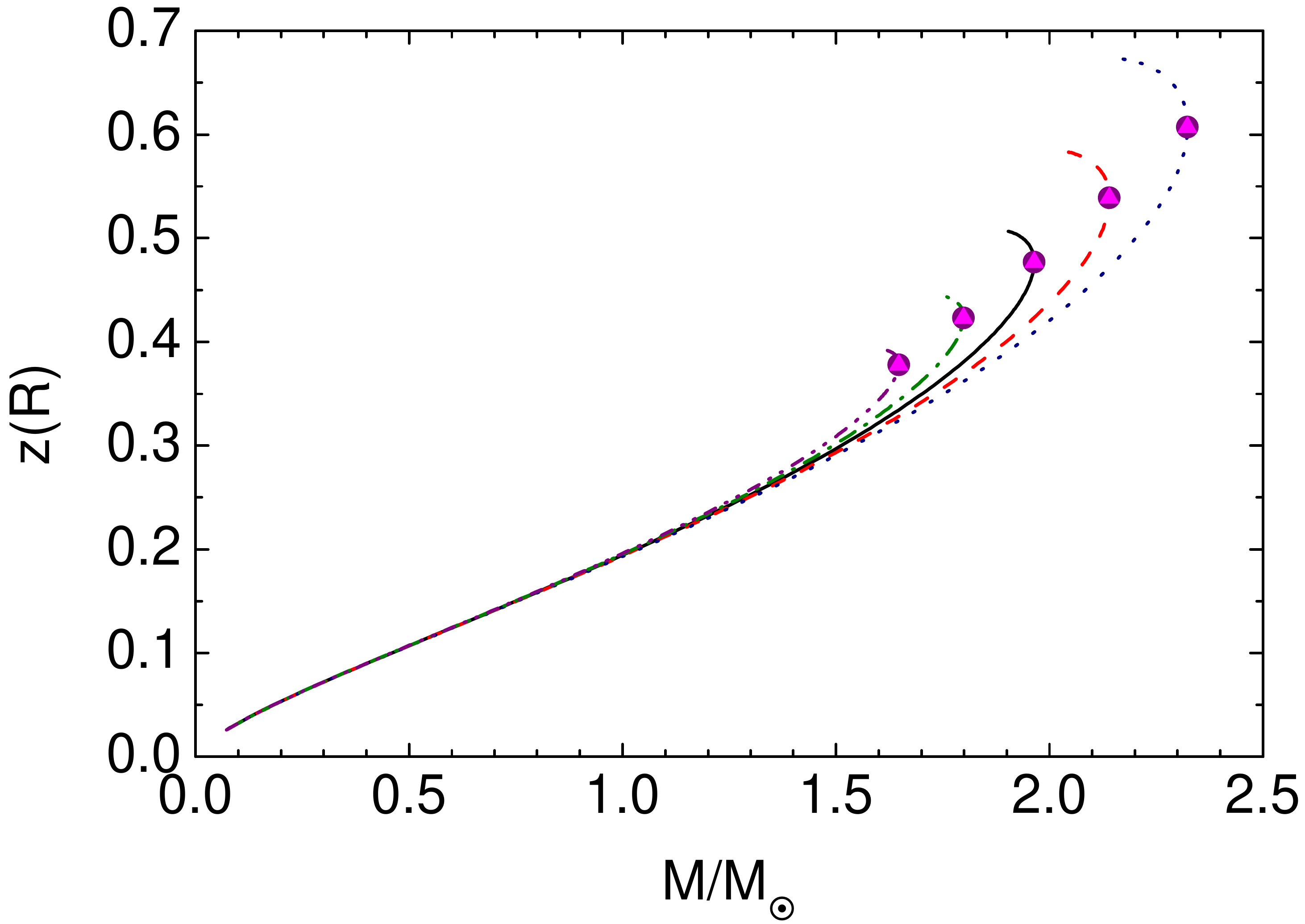}
\caption{\label{z_M_Carga0}Redshift function $z(R)=e^{-\nu(R)/2}-1$ at star's surface against the total mass for few values of $\kappa$. On the left panel is employed the AF $1$ and on the right panel is used the AF $2$. On the curves, the full circles and the full triangles mark the maximum mass points and the places where the zero oscillation frequencies are attained.}
\end{figure} 

A very important point that has to be analyzed is the redshift at the star's surface. The redshift indicates that the light emitted from an object is shifted toward the red end of the spectrum. Thus, the behavior of the redshift at the surface of an anisotropic strange star, $z(R)=e^{-\nu(R)/2}-1$ as a function of the total mass is analyzed in Fig.~\ref{z_M_Carga0} for different values of $\kappa$, considering the AF $1$ on the left hand side panel and the AF $2$ on the right hand side panel. We can observe that the redshift increases with the total mass until reaching a maximum mass values, which are marked by the full circles, and after this point the stars are unstable.

On the other hand, in Fig.~\ref{z_M_Carga0}, it can also be noted that the redshift changes with $\kappa$. If we compare the results obtained in anisotropic case with those found in the isotropic case, we see that the anisotropy in a fluid produces changes in the value of the redshift. Provided that $\kappa>0$, for a greater value of $\kappa$ a larger value of the redshift is obtained. In other words, when the anisotropy $\sigma>0$, a larger $\sigma$ a larger $z(R)$ is found. This means that an infinite redshift might be possible, but only if the anisotropy $\sigma$ is infinite (see \cite{bowers_liang1974}).

\subsubsection{Stability of anisotropic strange stars with fixed $\kappa$}

\begin{figure}[tbp]
\centering
\includegraphics[width=0.45\linewidth,angle=0]{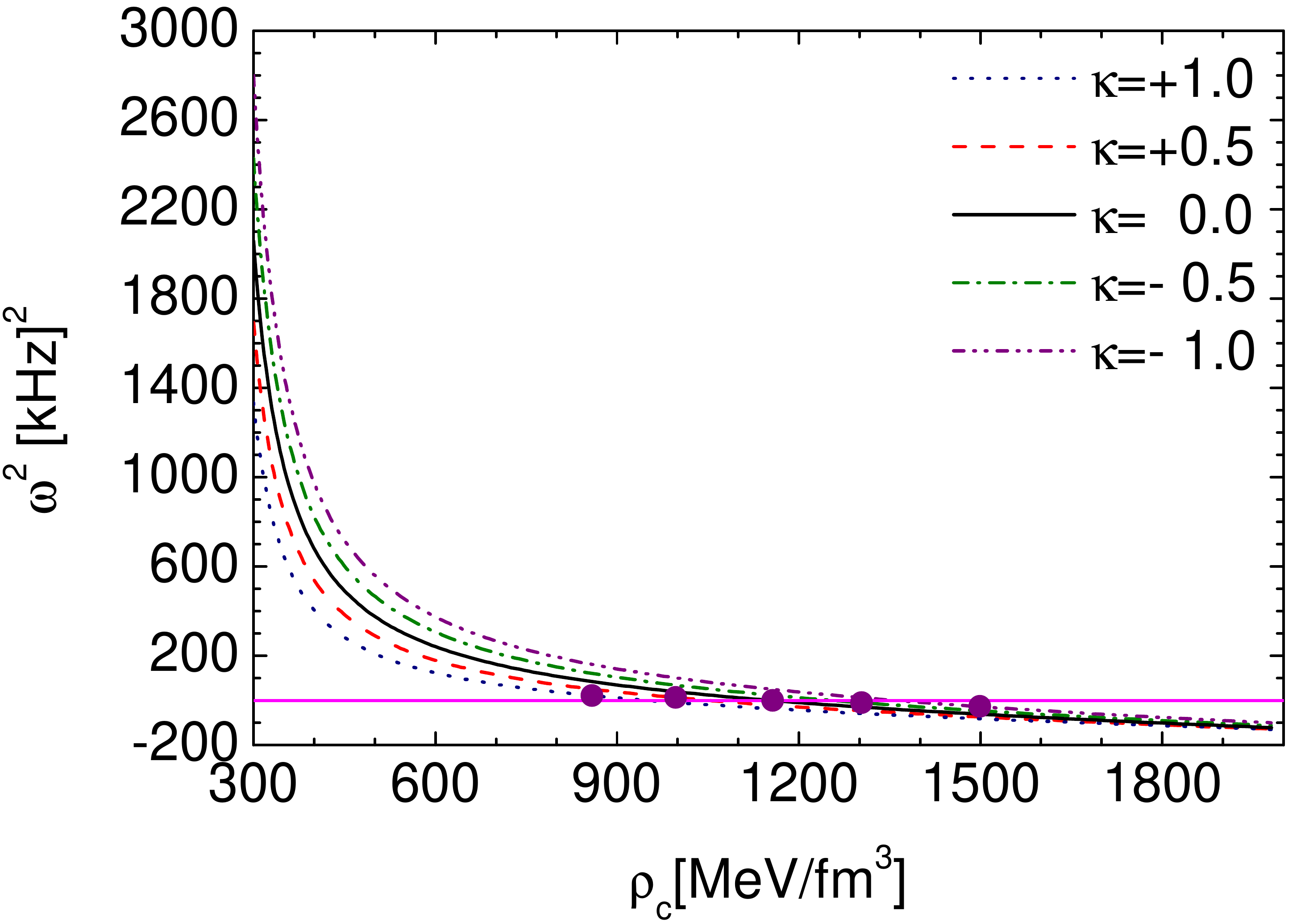}
\hfill
\includegraphics[width=0.45\linewidth,angle=0]{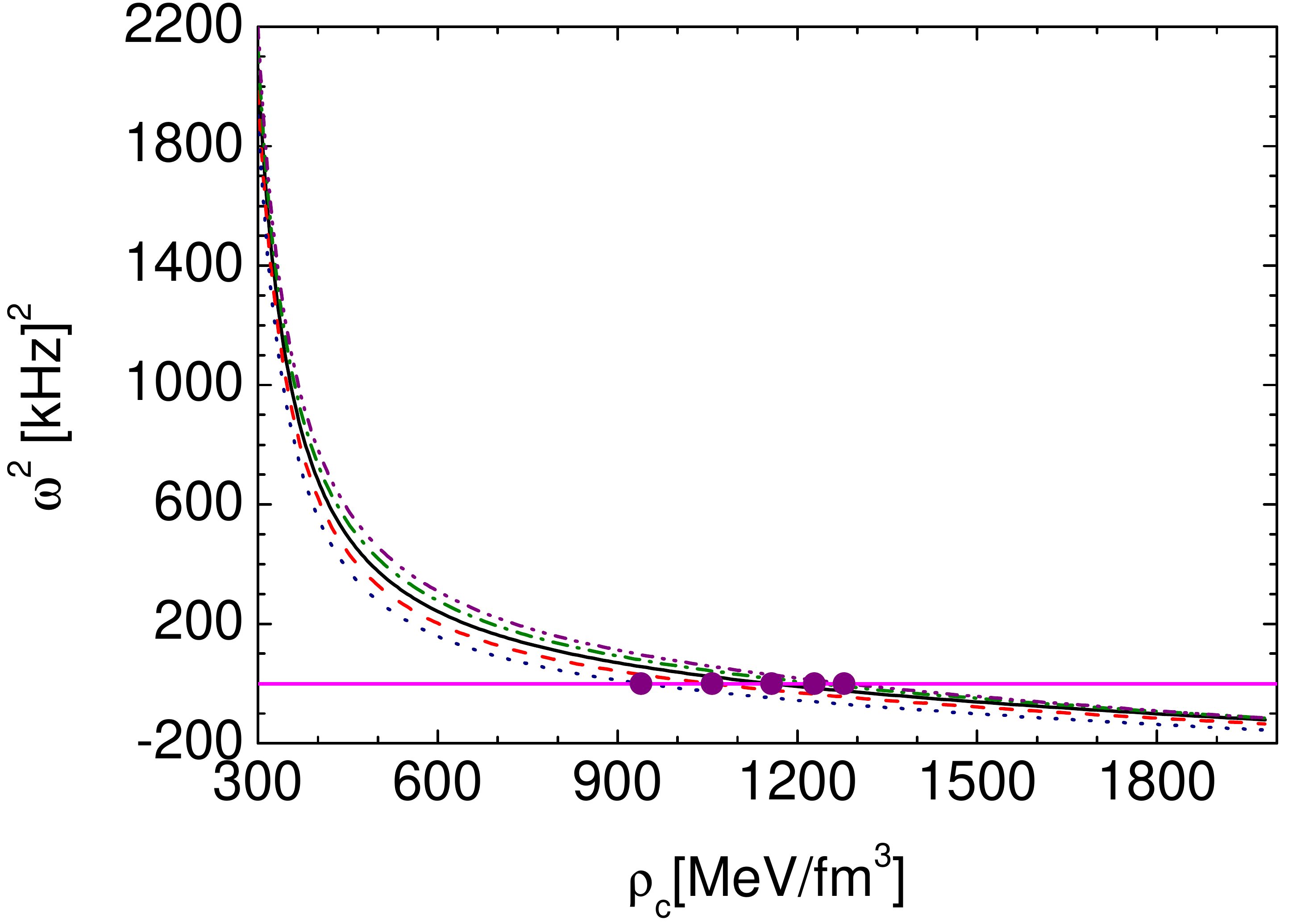}
\caption{\label{w_rhoc_Carga0}The frequencies of oscillation squared as function of the central energy density for few values of $\kappa$. On the left panel and on the right one are considered the AF $1$ and AF $2$, respectively. The full circles indicate the places where the maximum mass points are found and the horizontal line shows the places where the zero oscillation frequencies are attained.}
\end{figure}

The oscillation frequency of the fundamental mode squared as a function of the central energy density for different values of $\kappa$ is shown in Fig.~\ref{w_rhoc_Carga0}, using the AF $1$ on the left panel and employing the AF $2$ on the right panel. The spheres on the curves represent the points where the maximum mass values are reached and the horizontal line indicates the places where the oscillation frequencies are zero $\omega=0$. In all cases analyzed, we note that the oscillation frequency of the fundamental mode decreases with $\rho_c$, thus indicating that an object with larger central energy density will have lower stability. We also note that the anisotropy affects the stability of the star. For both cases $1$ and $2$, for a range of central energy density, since $\kappa>0$ ($\kappa<0$), the increment (reduction) of $\kappa$ decreases (grows) the stability of the star. 

In addition, it is important to mention that in Fig.~\ref{w_rhoc_Carga0}, in the case $1$, where $\sigma_s\neq0$, as mentioned before, we clearly note that $M_{\rm max}/M_{\odot}$ and $\omega=0$ are determined in a different central energy density. In the case $2$, where $\sigma_s=0$, as aforesaid, $M_{\rm max}/M_{\odot}$ and $\omega=0$ are reached at the same $\rho_c$. 

\begin{figure}[tbp]
\centering
\includegraphics[width=0.45\linewidth,angle=0]{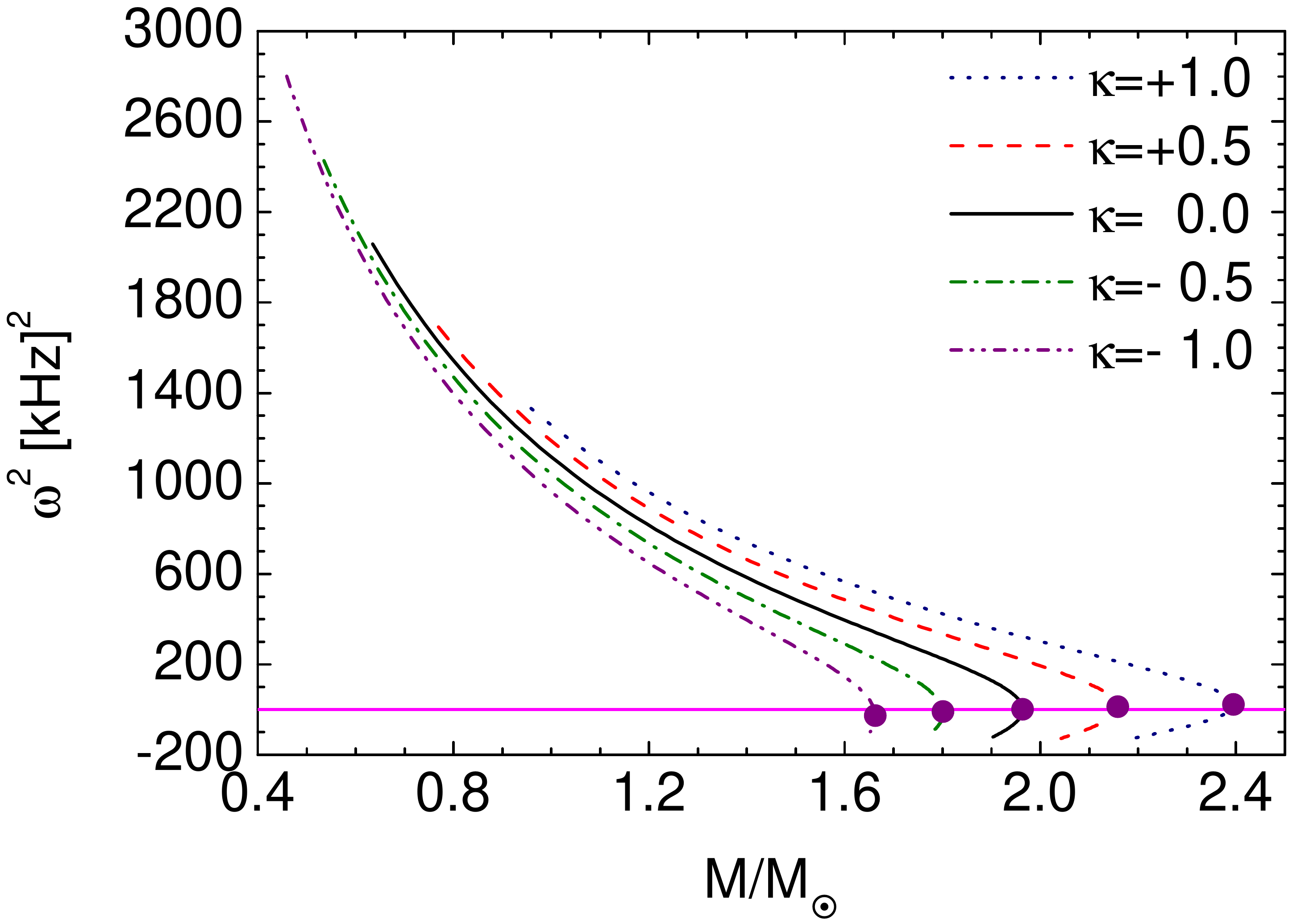}
\hfill
\includegraphics[width=0.45\linewidth,angle=0]{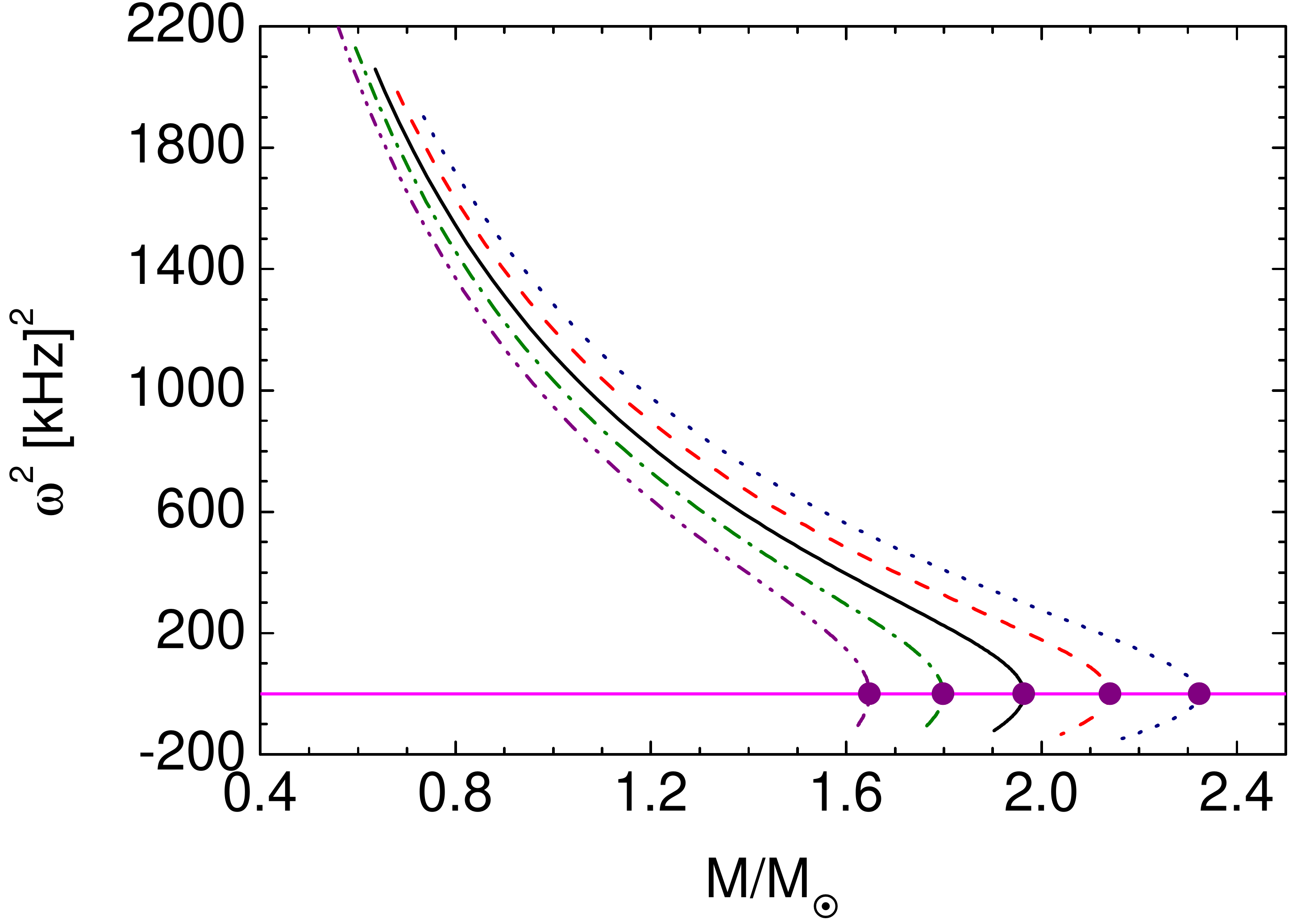}
\caption{\label{w_M_Carga0}The oscillation frequencies of the fundamental mode squared versus the total mass for five different values of $\kappa$. On the left panel is used the AF $1$ and on the right panel is employed the AF $2$. The full circles mark the places where the maximum mass points are attained and the horizontal line indicates the places where the zero oscillation frequencies are found.}
\end{figure}

The frequency of oscillation squared against the total mass for some values of $\kappa$ is presented in Fig.~\ref{w_M_Carga0}, considering the AF $1$ and AF $2$ respectively on the left and on the right. Independently of the value of $\kappa$ used, we obtain that $\omega^2$ decreases  with the increment of the mass. In the case $1$, when $\sigma_s\neq0$, the zero frequency of oscillation is not reached at the maximum mass point. In turn, in the case $2$, for $\sigma_s=0$, $\omega=0$ and $M_{\rm max}/M_{\odot}$ match at the same point indicating that the maximum mass point marks the onset of the instability. On the other hand, it can be seen in the curves that the anisotropy affects the stability of the object. Since $\kappa>0$, the increment of $\kappa$ helps to grow the stellar stability, in turn, once $\kappa<0$, the diminution of $\kappa$ brings as a consequence the reduction of the stellar stability.

\begin{table}[tbp] 
\centering
\begin{tabular}{|c|c|c|c|c|}
\hline
\multicolumn{5}{ |c| }{Case $1$}\\
\hline
$\kappa$   & $M_{\rm max}/M_{\odot}$& $R\,[\rm km]$& $\rho_c\,[\rm MeV/fm^3]$& $\omega^2\,[\rm kHz]^2$ \\ \hline
$+1.0$     & $2.39541$              & $11.6512$    & $858.521$               & $+22.0096$                  \\
$+0.5$     & $2.15849$              & $11.1423$    & $1002.62$               & $+11.5595$                  \\
\;\;\;$0.0$& $1.96405$              & $10.7120$    & $1155.01$               & $0.0$                  \\
$-0.5$     & $1.80119$              & $10.3278$    & $1332.50$               & $-15.5156$                  \\
$-1.0$     & $1.66259$              & $10.0042$    & $1502.31$               & $-27.8664$                  \\
\hline
\hline
$\kappa$   & $M/M_{\odot}$          & $R\,[\rm km]$& $\rho_c\,[\rm MeV/fm^3]$& $\omega^2\,[\rm kHz]^2$ \\ \hline
$+1.0$     & $2.39037$              & $11.4873$    & $952.942$               & $0.0$                  \\
$+0.5$     & $2.15754$              & $11.0762$    & $1053.63$               & $0.0$                  \\
\;\;\;$0.0$& $1.96405$              & $10.7120$    & $1155.01$               & $0.0$                  \\
$-0.5$     & $1.80063$              & $10.3859$    & $1257.35$               & $0.0$                  \\
$-1.0$     & $1.66074$              & $10.0913$    & $1360.78$               & $0.0$                  \\
\hline
\end{tabular}
\caption{\label{table1}Top: the constant $\kappa$ used and the stellar configurations with maximum masses with their respective total radii, central energy densities, and the frequencies of oscillation of the stars. Bottom: the constant $\kappa$ employed and the equilibrium configurations where the zero frequencies of oscillation are found. In both tables, the AF $1$ is used.}
\end{table}

Some anisotropic equilibrium configurations for the case $1$, $\sigma_s\neq0$, are presented in Table~\ref{table1}. On the top, the constants $\kappa$ used and the maximum masses with their respective total radii, central energy densities, and the frequencies of oscillation are shown, in turn, on the bottom, the constants $\kappa$ employed and the total masses with their respective total radii, central energy densities with zero oscillation frequencies are presented. In Table~\ref{table1}, we clearly note that the maximum masses and the zero oscillation frequencies are found using a different central energy density. 

\begin{table}[tbp] 
\centering
\begin{tabular}{|c|c|c|c|c|}
\hline
\multicolumn{5}{ |c| }{Case $2$}\\
\hline
$\kappa$   & $M_{\rm max}/M_{\odot}$& $R\,[\rm km]$& $\rho_c\,[\rm MeV/fm^3]$& $\omega^2\,[\rm kHz]^2$ \\ \hline
$+1.0$     & $2.32344$              & $11.1936$    & $941.707$              & $0.0$                   \\
$+0.5$     & $2.14033$              & $10.9456$    & $1055.23$              & $0.0$                   \\
\;\;\;$0.0$& $1.96405$              & $10.7120$    & $1155.01$              & $0.0$                  \\
$-0.5$     & $1.79878$              & $10.4910$    & $1231.15$              & $0.0$                  \\
$-1.0$     & $1.64741$              & $10.2799$    & $1280.47$              & $0.0$                  \\
\hline
\end{tabular}
\caption{\label{table2}The constants $\kappa$ considered and the stellar configurations with maximum masses with their respective total 
radii, central energy densities, and the frequencies of oscillations. The AF $2$ is considered.}
\end{table}

The values used for $\kappa$ and the equilibrium configurations with maximum masses with their respective total radii, central energy densities, and the frequencies of oscillation, for the case $2$, $\sigma_s=0$, are shown in Table~\ref{table2}. In this case, as aforesaid, the maximum masses and the zero frequency of oscillations are obtained at the same central energy density. 


\subsection{Equilibrium and stability of strange stars with fixed $\sigma_s$}

In case $1$ (where $\sigma_s\neq0$), unlike the case $2$ (where $\sigma_s=0$), the maximum mass point and the zero frequency of oscillation are found at different central energy densities. 
This drawback can be solved by using the equilibrium configurations found with different values of $\kappa$, to build other sequence of equilibrium configurations with the same value of $\sigma_s$. 

\subsubsection{Equilibrium of anisotropic strange stars with fixed $\sigma_s$}

\begin{figure}[tbp]
\centering
\includegraphics[width=0.32\linewidth,angle=0]{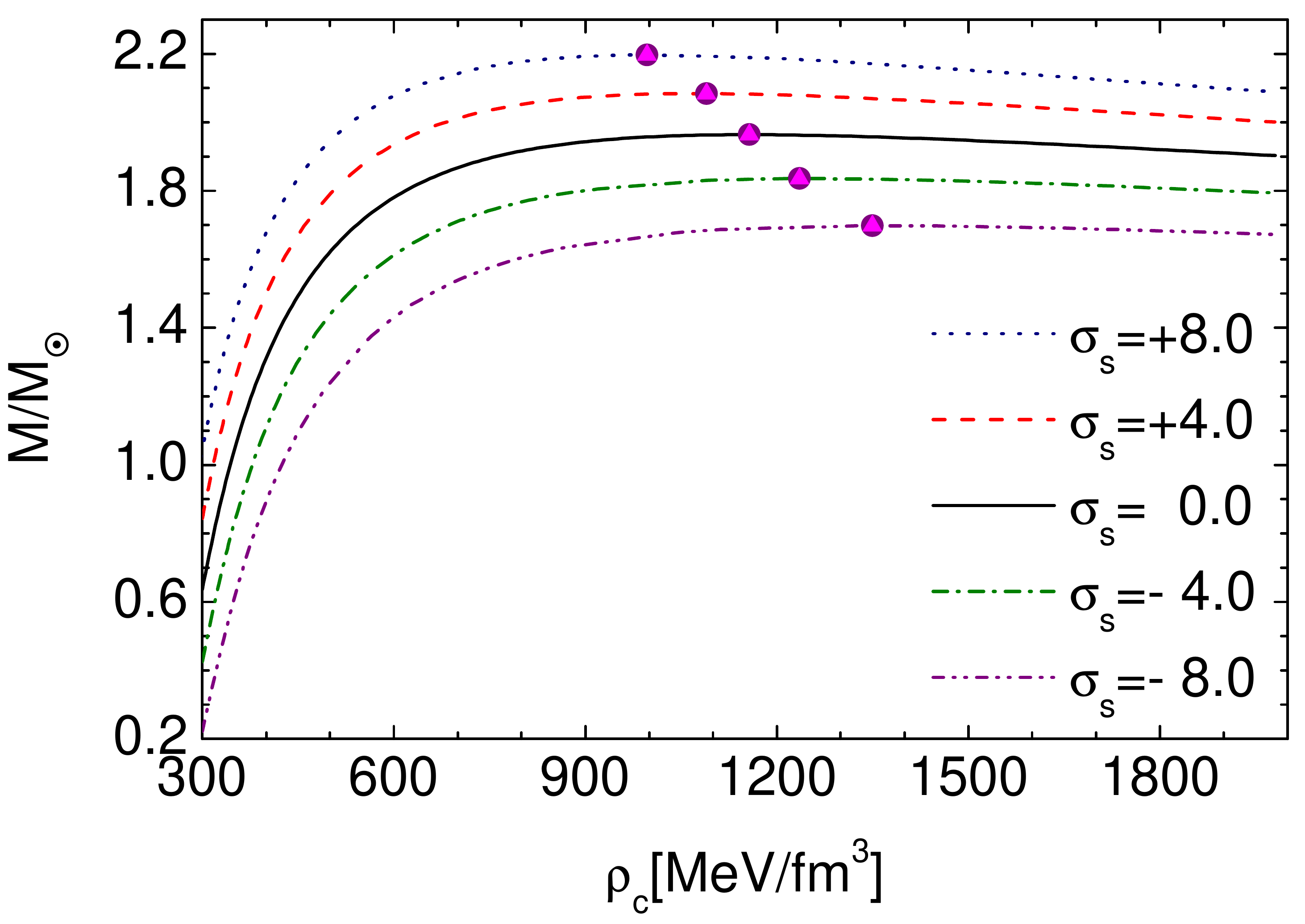}
\hfill
\includegraphics[width=0.32\linewidth,angle=0]{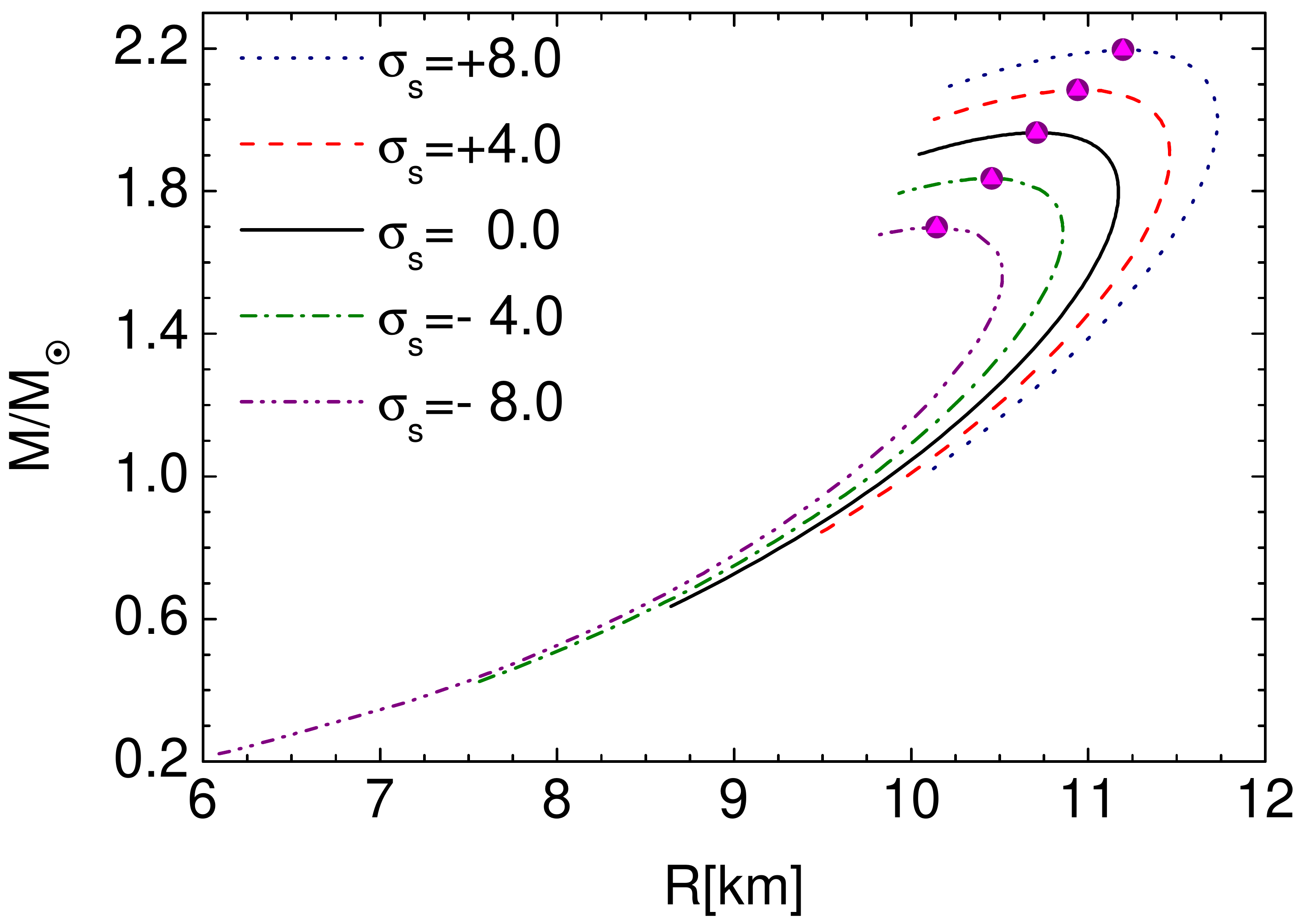}
\hfill
\includegraphics[width=0.32\linewidth,angle=0]{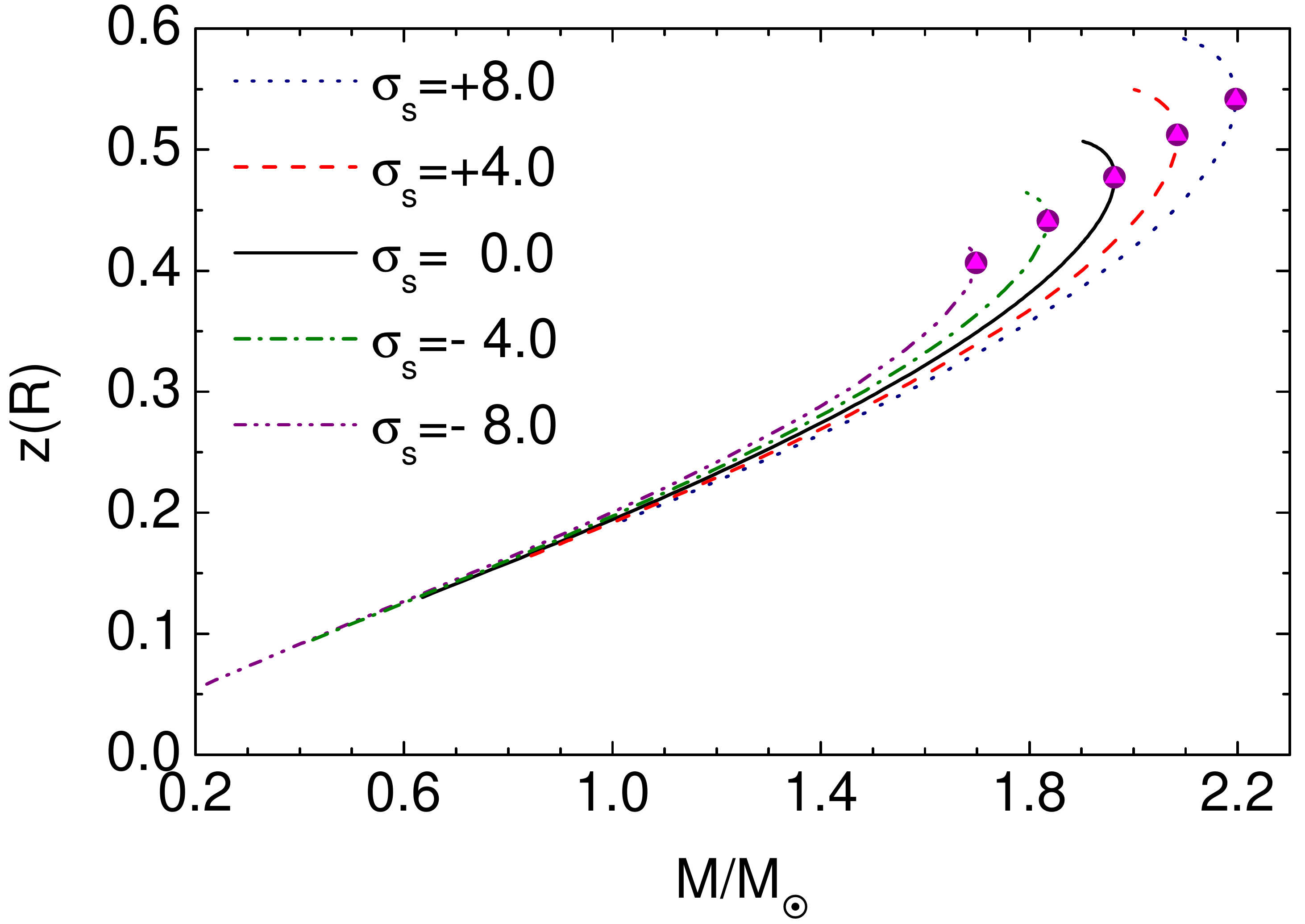}
\caption{\label{m_rho_sigma}On the left panel is shown the total mass of the star $M/M_{\odot}$ as a function of the central energy density, while in the middle panel is presented the behavior of the total mass versus of the total radius of the star, and on the right panel is displayed the mass against the redshift on the surface of the star. In the three panels, as indicated, are used five different values of $\sigma_s$. The units of $\sigma_s$ are ${\rm MeV/fm^3}$. The full circles and the full triangles indicate the places where the maximum mass points and the zero oscillation frequencies are attained, respectively.}
\end{figure}

In Fig.~\ref{m_rho_sigma}, the behavior of the total mass with the central energy density is presented on the left panel, whereas in the middle panel the total mass against the total radius of the star is shown, and on the right panel the redshift on the star surface versus the total mass of the star is plotted. In all figures are used five different values of $\sigma_s$ and the AF $1$. We can note that the maximum mass points (full circles) and the zero oscillation frequencies (full triangles) overlap only when is built a sequence of stars with the same value $\sigma_s$. This indicates that in a sequence of an equilibrium configurations with $\sigma_s$ fixed, the conditions $dM/d\rho_c>0$ and $dM/d\rho_c<0$ can always be used to recognize the stable regions from the unstable ones.

\subsubsection{Stability of anisotropic strange stars with fixed $\sigma_s$}

\begin{figure}[tbp]
\centering
\includegraphics[width=0.45\linewidth,angle=0]{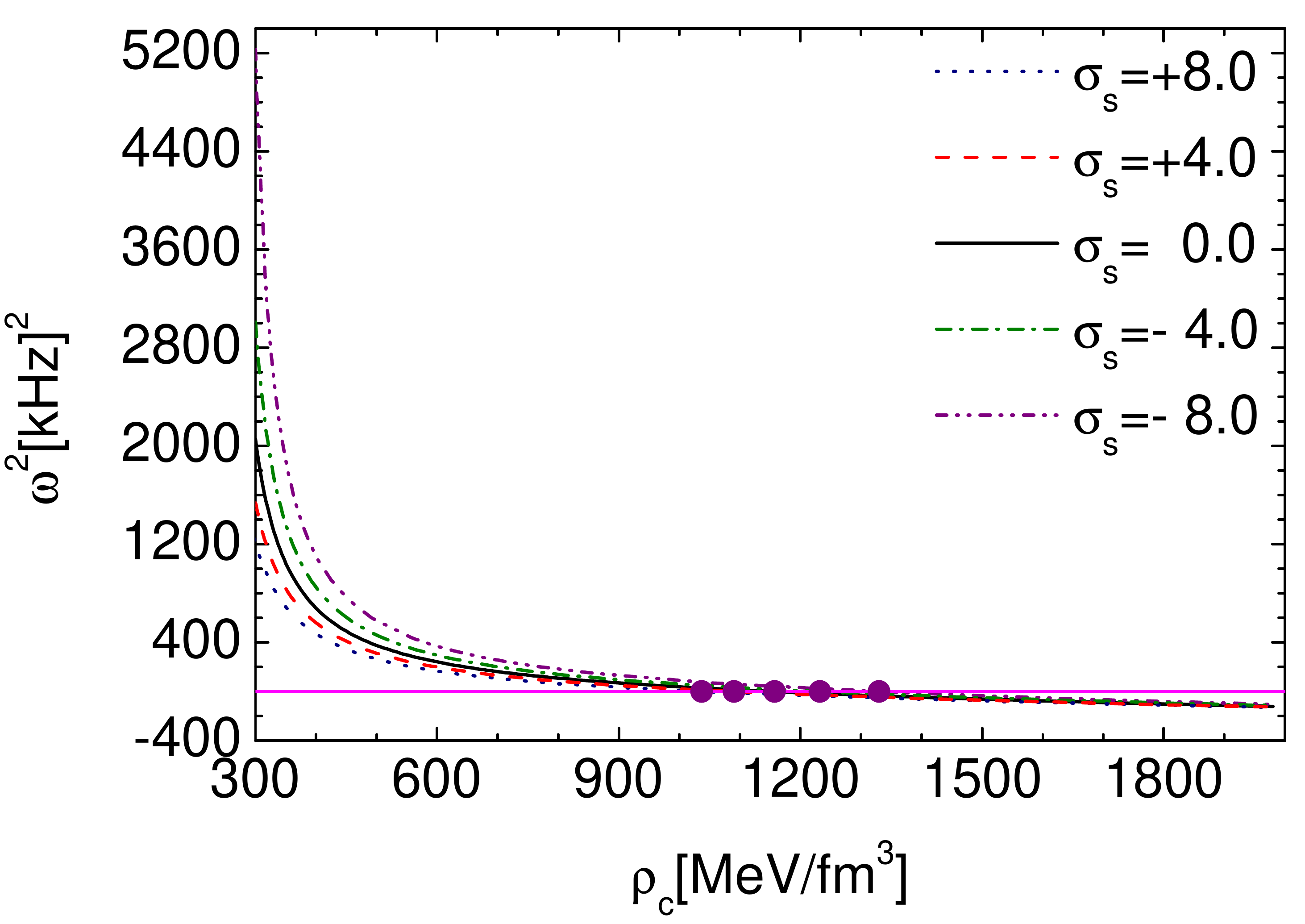}
\hfill
\includegraphics[width=0.45\linewidth,angle=0]{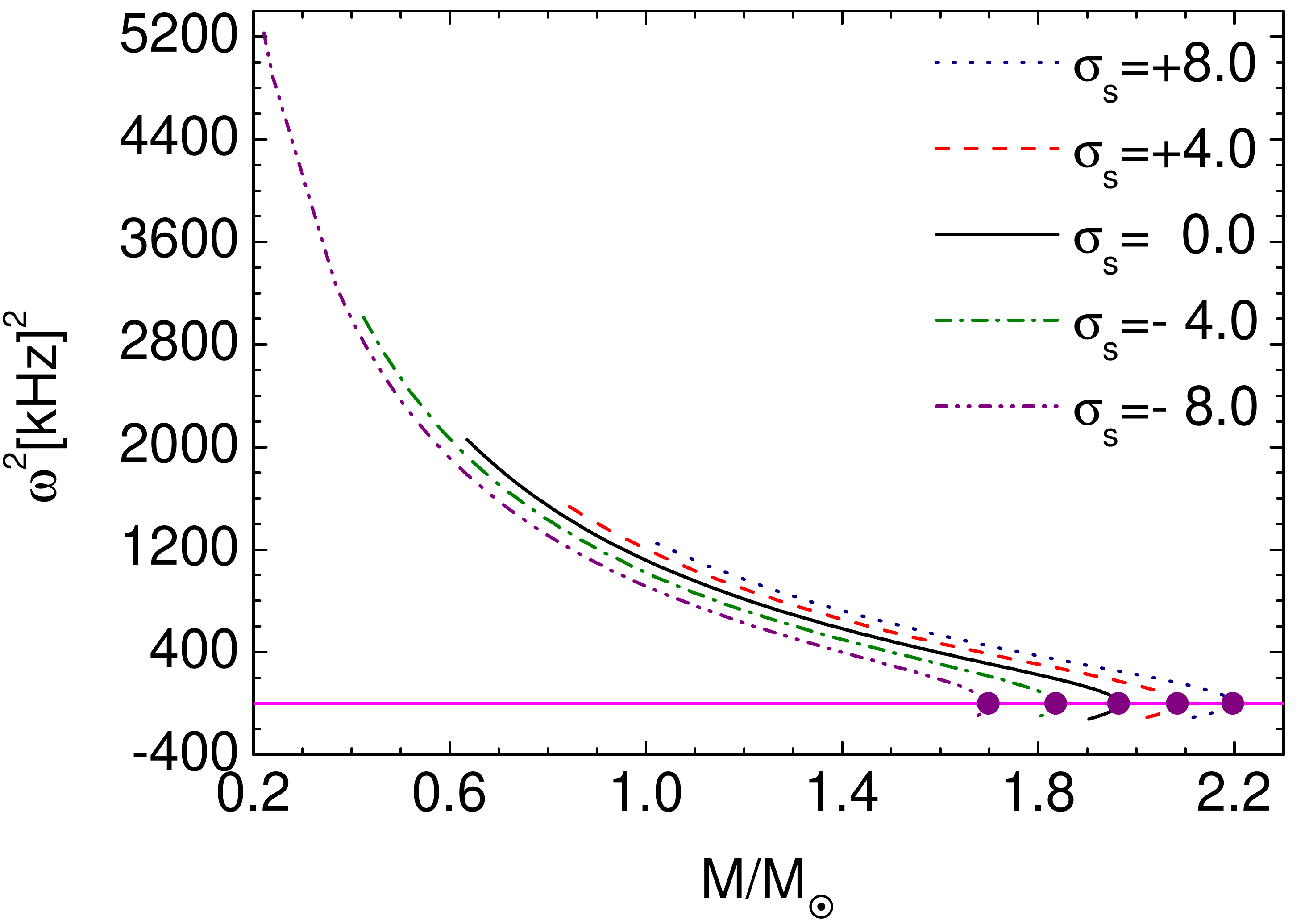}
\caption{\label{w2_m_sigma}The behavior of the oscillation frequencies of the fundamental mode squared with the central energy density and with the total mass for five different values of $\sigma_s$ are presented on the left panel and on the right panel, respectively. The units of $\sigma_s$ are ${\rm MeV/fm^3}$. The full circles mark the maximum mass points and the horizontal line shows the place where the zero frequency of oscillations are reached.}
\end{figure}

The frequency of oscillation of the fundamental mode squared against the central energy density and as a function of the total mass are presented on the left and on the right panels of Fig.~\ref{w2_m_sigma}, respectively, for five different values of $\sigma_s$. On the left panel, in all cases, we see that $\omega^2$ decreases with the increment of the central energy density, indicating that for larger central energy density lower stability is obtained. We also see that for a range of values of $\rho_c$ the stability is affected by the change of $\sigma_s$. Since $\sigma_s>0$, the increment of the anisotropy at the star surface $\sigma_s$ decreases the stability of the star. On the right panel we note that $\omega^2$ decreases with the growth of $M/M_{\odot}$, attaining $\omega=0$ in the maximum mass point $M_{\rm max}/M_{\odot}$. We also note that for a range of masses the value of $\sigma_s$ influences in the stability of the star: when $\sigma_s>0$,  the increment of $\sigma_s$ increases the stability of the star.

\begin{table}[tbp] 
\centering
\begin{tabular}{|c|c|c|c|c|}
\hline
\multicolumn{5}{ |c| }{Case $1$}\\
\hline
$\sigma_s\,[\rm MeV/fm^3]$   & $M_{\rm max}/M_{\odot}$& $R\,[\rm km]$& $\rho_c\,[\rm MeV/fm^3]$& $\omega^2\,[\rm kHz]^2$ \\ \hline
$+8.0$                       & $2.19600$              & $11.1459$    & $1035.47$               & $0.0$                   \\
$+4.0$                       & $2.08406$              & $10.9405$    & $1089.95$               & $0.0$                  \\
\;\;\;$0.0$                  & $1.96405$              & $10.7120$    & $1155.01$               & $0.0$                  \\
$-4.0$                       & $1.83562$              & $10.4574$    & $1233.74$               & $0.0$                  \\
$-8.0$                       & $1.69798$              & $10.1712$    & $1331.53$               & $0.0$                  \\
\hline
\end{tabular}
\caption{\label{table3}The constants $\sigma_s$ considered and the equilibrium configurations with maximum total masses with their respective total radius, central energy density, and the frequencies of oscillations. The AF $1$ is used.}
\end{table}

In Table~\ref{table3}, the values used for $\sigma_s$ and the maximum mass values with their respective total radii, central energy densities and frequency of oscillations are shown. From the values presented in the table, we can check that the maximum mass point and the zero frequency of oscillation are determined at the same central energy density, when $\sigma_s$ is fixed. This indicates that the maximum mass point marks the onset of the instability.

\section{Conclusions}\label{conclusion}

The radial stability of anisotropic strange quark stars is studied in the present article. This is possible through the numerical solution of the hydrostatic equilibrium equation and the radial pulsation equation, which are modified from their original form to the inclusion of anisotropy. We consider that the matter contained in the strange stars follows the MIT bag model equation of state. On the other hand, two different kinds of anisotropic factor $\sigma$ are considered to describe the anisotropy. One that is nonvanishing on the surface of the star and other one that vanishes on it, namely,  $\sigma_s\neq0$ (in case $1$) and $\sigma_s=0$ (in case $2$), respectively. Specifically, the anisotropic factors considered take the form $\sigma=\frac{\kappa}{1.7}(\rho+p_r)(\rho+3\,p)r^2 e^{\lambda}$, for the case $1$, and $\sigma=\kappa p_r (1-e^{-\lambda})$, for the case $2$. 

In the two cases analyzed, we found that the anisotropy affects physical properties of the stars, such as: the energy density, the radial pressure, the total radius, the total mass, the surface redshift, and the frequency of oscillation of the fundamental mode. In case $1$, when $\sigma_s\neq0$, we found that the maximum mass point and the zero frequency of oscillation are obtained for different central energy densities. From this we understand that the conditions $dM/d\rho_c>0$ and $dM/d\rho_c<0$ are necessary but not sufficient to recognize regions corresponding to stable and unstable stars in a sequence of equilibrium configurations with fixed $\kappa$. In turn, in the case $2$, when $\sigma_s=0$, the maximum mass point and $\omega=0$ are found at the same central energy density. This indicates that the maximum mass point marks the onset of the instability in this case. In other words, when $\sigma_s=0\ (p_t(R)=p_r(R)=0)$, the conditions $dM/d\rho_c>0$ and $dM/d\rho_c<0$ can be used to distinguish between a stable and an unstable region in a sequence of equilibrium configurations with fixed $\kappa$.

In case $1$ (when $\sigma_s\neq0$), the maximum mass point and the zero frequency of oscillation are obtained at the same central energy density only in a system of equilibrium configurations with the same $\sigma_s \ (p_t(R)\neq 0\ \textrm{and fixed})$. This indicates that in such sequences with $\sigma_s$ fixed, the conditions $dM/d\rho_c>0$ and $dM/d\rho_c<0$ can be used to determine the regions constituted by stable or unstable against radial perturbations, respectively. In addition, we determine that for a range of some parameters, the anisotropy $\sigma_s$ influences the stability of the star. For a range of $\rho_c$, when the anisotropy at star surface $\sigma_s> 0$, the growth of $\sigma_s$ reduces the stability of the star. Moreover, for a range of masses, while $\sigma_s>0$,  the increment of $\sigma_s$ increases the stability of the star.

\

\acknowledgments
The authors thank Edson Otoniel, Brett Carlson and Pedro Moraes for reading the manuscript and for valuable suggestions. J.D.V.A. thanks Coordena\c{c}\~ao de Aperfei\c{c}oamento de Pessoal de N\'\i vel Superior - CAPES, Brazil, for a grant. M. M. would like to thank Funda\c{c}\~ao de Amparo \`a Pesquisa do S\~ao Paulo for financial support under the Thematic Project $13/26258-4$. 

\end{document}